\documentclass[twocolumn]{aastex61}
\pdfoutput=1

\begin{document}
%% Reintroduced the \received and \accepted commands from AASTeX v5.2
\received{2018}
\revised{2018}
\accepted{2018}
%% Command to document which AAS Journal the manuscript was submitted to.
%% Adds "Submitted to " the arguement.
%\submitjournal{ApJ}

%% Mark up commands to limit the number of authors on the front page.
%% Note that in AASTeX v6.1 a \collaboration call (see below) counts as
%% an author in this case.
%
%\AuthorCollaborationLimit=3
%
%% Will only show Schwarz, Muench and "the AAS Journals Data Scientist 
%% collaboration" on the front page of this example manuscript.
%%
%% Note that all of the author will be shown in the published article.
%% This feature is meant to be used prior to acceptance to make the
%% front end of a long author article more manageable. Please do not use
%% this functionality for manuscripts with less than 20 authors. Conversely,
%% please do use this when the number of authors exceeds 40.
%%
%% Use \allauthors at the manuscript end to show the full author list.
%% This command should only be used with \AuthorCollaborationLimit is used.

%% The following command can be used to set the latex table counters.  It
%% is needed in this document because it uses a mix of latex tabular and
%% AASTeX deluxetables.  In general it should not be needed.
%\setcounter{table}{1}

%%%%%%%%%%%%%%%%%%%%%%%%%%%%%%%%%%%%%%%%%%%%%%%%%%%%%%%%%%%%%%%%%%%%%%%%%%%%%%%%
%%
%% The following section outlines numerous optional output that
%% can be displayed in the front matter or as running meta-data.
%%
%% If you wish, you may supply running head information, although
%% this information may be modified by the editorial offices.
\shorttitle{The nuclear obscuring disk in the Compton-thick
  Seyfert galaxy NGC~5643 }
\shortauthors{Alonso-Herrero et al.}
%%
%% You can add a light gray and diagonal water-mark to the first page 
%% with this command:
% \watermark{text}
%% where "text", e.g. DRAFT, is the text to appear.  If the text is 
%% long you can control the water-mark size with:
%  \setwatermarkfontsize{dimension}
%% where dimension is any recognized LaTeX dimension, e.g. pt, in, etc.
%%
%%%%%%%%%%%%%%%%%%%%%%%%%%%%%%%%%%%%%%%%%%%%%%%%%%%%%%%%%%%%%%%%%%%%%%%%%%%%%%%%

%% This is the end of the preamble.  Indicate the beginning of the
%% manuscript itself with \begin{document}.

\title{Resolving the nuclear obscuring disk in the Compton-thick
  Seyfert galaxy NGC~5643 with ALMA}

\correspondingauthor{Almudena Alonso-Herrero}
\email{aalonso@cab.inta-csic.es}

\author{A. Alonso-Herrero}
\affil{Centro de Astrobiolog\'{\i}a (CAB, CSIC-INTA), ESAC Campus,
  E-28692 Villanueva de la Ca\~nada, Madrid, Spain}
\affiliation{Department of Physics, University of Oxford, Keble Road,
  Oxford OX1 3RH, United Kingdom}

\author{M. Pereira-Santaella}
\affiliation{Department of Physics, University of Oxford, Keble Road,
  Oxford OX1 3RH, United Kingdom}

\author{S. Garc\'{\i}a-Burillo}
\affiliation{Observatorio de Madrid, OAN-IGN, Alfonso XII, 3, E-28014
  Madrid, Spain}

\author{R. I. Davies}
\affiliation{Max Planck Institut fuer extraterrestrische Physik
Postfach 1312, 85741 Garching bei Muenchen, Germany}

\author{F. Combes}
\affiliation{LERMA, Obs. de Paris, PSL Research Univ., Coll\'ege de France, CNRS, Sorbonne Univ.,
  UPMC, Paris, France}
\author{D. Asmus}
\affiliation{European  Southern Observatory, Casilla 19001, Santiago
19, Chile}
\affiliation{Department of Physics \& Astronomy, University of Southampton, Hampshire SO17 1BJ, Southampton, United Kingdom}
\author{A. Bunker}
\affiliation{Department of Physics, University of Oxford, Keble Road,
  Oxford OX1 3RH, United Kingdom}
\author{T. D\'{\i}az-Santos}
\affiliation{N\'ucleo de Astronom\'{\i}a de la Facultad de
  Ingenier\'{\i}a, Universidad Diego Portales, Av. Ej\'ercito
  Libertador 441, Santiago, Chile}
\author{P. Gandhi}
\affiliation{Department of Physics \& Astronomy, University of Southampton, Hampshire SO17 1BJ, Southampton, United Kingdom}
\author{O. Gonz\'alez-Mart\'{\i}n}
\affiliation{Instituto de Radioastronom\'ia y Astrof\'isica (IRyA-UNAM), 3-72 (Xangari), 8701, Morelia, Mexico}
\author{A. Hern\'an-Caballero}
\affiliation{Departamento de Astrof\'{\i}sica y CC. de la Atm\'osfera, Facultad de CC. F\'{\i}sicas, Universidad Complutense de Madrid, E-28040 Madrid, Spain}
\author{E. Hicks}
\affiliation{Department of Physics and Astronomy, University of Alaska Anchorage, AK 99508-4664, USA}
\author{S. H\"onig}
\affiliation{Department of Physics \& Astronomy, University of Southampton, Hampshire SO17 1BJ, Southampton, United Kingdom}
\author{A. Labiano}
\affiliation{Centro de Astrobiolog\'{\i}a (CAB, CSIC-INTA), Carretera
  de Torrej\'on a Ajalvir, E-28850 Torrej\'on de Ardoz, Madrid, Spain}
\author{N. A. Levenson}
\affiliation{Space Telescope Science Institute, 3700 San Martin Drive,
  Baltimore, MD 21218, USA}
\author{C. Packham}
\affiliation{Department of Physics and Astronomy, University of Texas
  at San Antonio, 1 UTSA Circle, San Antonio, TX 78249, USA}
\author{C. Ramos Almeida}
\affiliation{Instituto de Astrof\'{\i}sica de Canarias, Calle v\'{\i}a L\'actea, s/n, E-38205 La Laguna, Tenerife, Spain}
\affiliation{Departamento de Astrof\'{\i}sica, Universidad de La Laguna, E-38205 La Laguna, Tenerife, Spain}
\author{C. Ricci}
\affiliation{N\'ucleo de Astronom\'ia de la Facultad de Ingenier\'ia,
  Universidad Diego Portales, Av. Ej\'ercito Libertador 441, Santiago,
  Chile}
\affiliation{Chinese Academy of Sciences South America Center for
  Astronomy, Camino El Observatorio 1515, Las Condes, Santiago, Chile}
 \affiliation{Kavli Institute for Astronomy and Astrophysics, Peking
   University, Beijing 100871, China}
\author{D. Rigopoulou}
\affiliation{Department of Physics, University of Oxford, Keble Road,
  Oxford OX1 3RH, United Kingdom}
\author{D. Rosario}
\affiliation{Centre for Extragalactic Astronomy, Durham University,
  South Road, Durham DH1 3LE, United Kingdom}
\author{E. Sani}
\affiliation{European  Southern Observatory, Casilla 19001, Santiago 19, Chile}
\author{M. J. Ward}
\affiliation{Centre for Extragalactic Astronomy, Durham University,
  South Road, Durham DH1 3LE, United Kingdom}



\begin{abstract}

  We present ALMA Band 6 $^{12}$CO(2--1) line and rest-frame 232\,GHz continuum
  observations of the nearby
  Compton-thick Seyfert galaxy NGC~5643 with angular resolutions
  $0.11-0.26\arcsec$ \, ($9-21\,$pc). The CO(2--1)
  integrated line map reveals  emission from the nuclear and circumnuclear
  region with a two-arm nuclear spiral extending
  $\sim 10\arcsec$ \, on each side. The circumnuclear CO(2--1) kinematics
  can be fitted with a rotating disk, although  there are regions 
with large residual
  velocities  and/or velocity dispersions.
  The CO(2--1) line profiles of these regions show two different velocity components.
  One is ascribed to the circular component and the other to
  the interaction of the AGN outflow, as traced by the [O\,{\sc
  iii}]$\lambda$5007\AA \, emission, with molecular
  gas in the disk a few hundred parsecs from the AGN. On nuclear scales, we detected an
  inclined CO(2--1) disk (diameter 26\,pc, FWHM) oriented almost in a
  north-south direction. 
The CO(2--1) nuclear kinematics can be fitted with a rotating disk which 
  appears to be tilted with respect to the
  large scale disk. There are strong non-circular motions in the
  central $0.2-0.3\arcsec$ with velocities of up to $110\,{\rm
    km\,s}^{-1}$. In the absence of a nuclear bar, these motions could be explained as
  radial outflows in the nuclear disk.
We estimate
  a total molecular gas mass for the  
  nuclear disk of $M({\rm H}_2)=1.1\times 10^7\,M_\odot$ and an
  H$_2$ column density toward the
  location of the AGN of $N({\rm H}_2)\sim 5 \times 10^{23}\,{\rm
    cm}^{-2}$, for a standard
  CO-to-H$_2$ conversion factor. 
  We interpret this nuclear molecular gas disk
  as the obscuring torus of NGC~5643 as well as the collimating structure of the
  ionization cone.

\end{abstract}




\section{Introduction} \label{sec:intro}

The key piece of the Unified Model for active galactic nuclei (AGN) is
the torus of dust and molecular gas that obscures the direct view of
the  AGN along certain lines of sight and explains
the observational properties of AGN \citep{Antonucci1993}. 
ALMA observations detected the torus in
the nearby Seyfert NGC~1068 in both cold dust and molecular line
emission \citep{GarciaBurillo2016, Gallimore2016, Imanishi2018}.  The measured torus
diameter  at different ALMA molecular transitions and dust
  continuum  ($7-13\,$pc) is approximately a factor of two larger than the size
derived from the modelling of the nuclear unresolved near-to-mid infrared
emission \citep{RamosAlmeida2011, AlonsoHerrero2011, Ichikawa2015} and
mid-infrared interferometry \citep{Tristram2009, Burtscher2013,
  LopezGonzaga2014}. Therefore, the full extent of the torus
is larger than that of the warm dust probed by the near and mid-infrared continuum.
Indeed, \cite{Fuller2016} showed that when including data to $\sim
40\,\mu$m the parameterization of the torus does 
indeed alter, further demonstrating the need for a broad wavelength sampling of the torus emission.

%This suggests that near to mid-infrared
%observations might  not be able to 
%probe the {\it full extent of the torus}, while ALMA is sensitive to the
%bulk of the torus through its cold dust and molecular gas emission.

Cold molecular gas is detected in the nuclear and circumnuclear
regions of nearby Seyfert galaxies on physical scales of tens to hundreds
of parsecs \citep[e.g.,][]{Tacconi1994, Schinnerer2000, 
  Krips2007, 
  GarciaBurillo2005, GarciaBurillo2014, Sani2012, Combes2013, Izumi2016, Lin2016,
  Zschaechner2016, Salak2017}. This gas is believed to be
associated with the AGN fuelling processes. 
Indeed, in local AGN the hot \citep[$T\sim 1000-2000\,$K, see][]{Mouri1994}
molecular gas,
as traced by the near-IR  rovibrational $2.12\,\mu$m 1--0 S(1) H$_2$ line, is
more centrally concentrated than in non active galaxies and believed to be
related to the Unified Model obscuring torus  \citep{Hicks2009, Hicks2013}.  

A natural consequence of the accumulation of material in the
central regions of active galaxies   is  the presence of nuclear 
($<100$\,pc) on-going/recent star formation  activity 
\citep[e.g.,][]{Davies2007, Esquej2014}. Since the
nuclear star formation rate
is found to be correlated with the velocity dispersion of the nuclear
hot molecular gas disks \citep{Hicks2013}, then the nuclear molecular gas
disks could be maintained by 
inflows of material into the nuclear region and/or by intense, short-lived
nuclear star formation through stellar feedback.

We have started several ALMA programs to observe the cold molecular gas in
the nuclear and circumnuclear regions of a hard X-ray selected
sample of 
nearby Seyfert galaxies drawn from
the X-ray {\it Swift}/BAT all sky 70 month catalog \citep{Baumgartner2013}.
The goal of these ALMA programs is to 
understand the connections between the cold and hot molecular gas, the AGN torus and nuclear/circumnuclear star formation
activity in AGN. As explained above, 
the near-infrared H$_2$  lines provide useful 
information about the morphology and kinematics of the hot molecular gas
in the nuclear regions of Seyfert galaxies. However, they only  probe
a small fraction of the total molecular gas fraction in galaxies
\citep[see e.g., ][]{Dale2005}.

In this work we present ALMA Band 6 continuum and $^{12}$CO(2--1) observations
of the nearby ($D=16.9\,$Mpc, 
$1\arcsec = 81.9\,$pc) Seyfert 2 galaxy NGC~5643. Although this galaxy
is classified as an SAB(rs)c, it has a large-scale stellar bar identified in
the near-infrared \citep{Mulchaey1997, Jungwiert1997}. It is a
Compton-thick \citep[see][]{Guainazzi2004, Ricci2015, Annuar2015}
galaxy with a moderate intrinsic X-ray luminosity ($L_{\rm 2-10keV} \sim
10^{42}\,{\rm erg\,s}^{-1}$). Optical broad lines in polarized light have not been detected in this
  galaxy \citep{RamosAlmeida2016}. These authors explained it as a combination of its Compton-thick 
  nature and relatively low AGN luminosity or
  different properties of the scattering material.

The $2.12\,\mu$m 1--0 S(1) H$_2$ hot molecular gas in the central region of NGC~5643
shows anomalous kinematics in an area  
at about $2\arcsec$ northeast of the nucleus with high velocity
dispersion \citep{Davies2014}. This  might be a signature of the presence of
molecular gas being excited near the edge of the ionization cone
traced by the optical
[O\,{\sc iii}]$\lambda$5007\AA\, line \citep{Simpson1997}. The radio emission of NGC~5643 shows an almost east-west
orientation extending for about 30\arcsec \, on both sides of the nucleus
\citep{Morris1985,
  Leipski2006}.  This  radio structure appears to be impacting on the
disk on the east side of the galaxy  producing positive feedback as
revealed by the presence of H\,{\sc ii} regions at approximately 5 and
10\arcsec \, east of the nucleus \citep{Cresci2015}. The soft X-ray emission
  is also extended and mostly detected in the east side of the galaxy
  following the [O\,{\sc iii}] emission \citep{Bianchi2006, GomezGuijarro2017}.

The paper is organized as follows. Section~\ref{sec:obs} describes
the ALMA observations as well as archival optical integral field
spectroscopy. In Sections~\ref{sec:circumnuclear}
and \ref{sec:nuclear} we present the analysis of the ALMA CO(2--1)
morphology and kinematics of the circumnuclear and nuclear region of
NGC~5643, respectively. In Section~\ref{sec:conclusions} we discuss
the results and give our conclusions. 

\newpage

\begin{figure*}[h]
\hspace{3.5cm}
\includegraphics[width=0.6\textwidth]{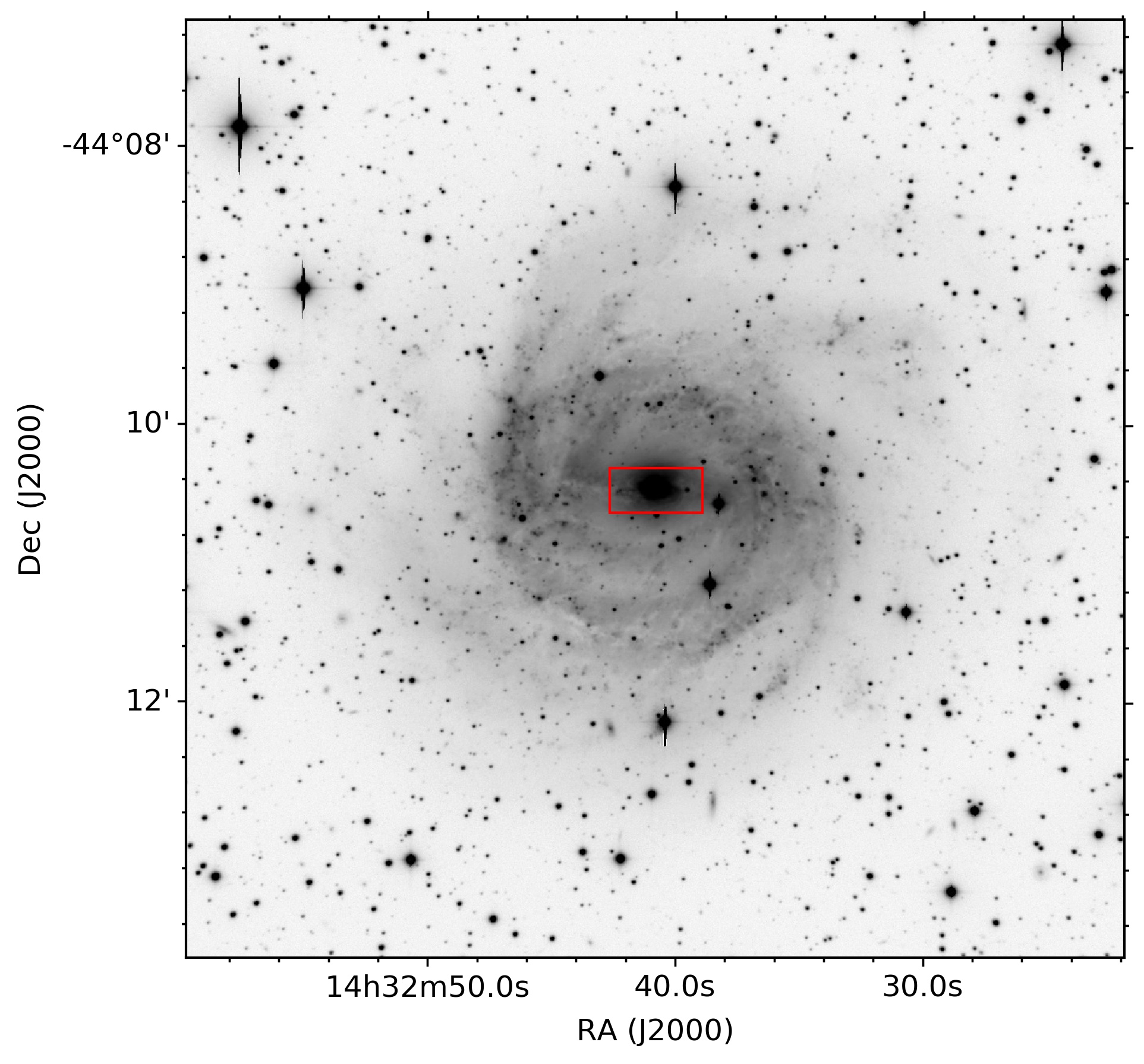}

\includegraphics[width=1.\textwidth]{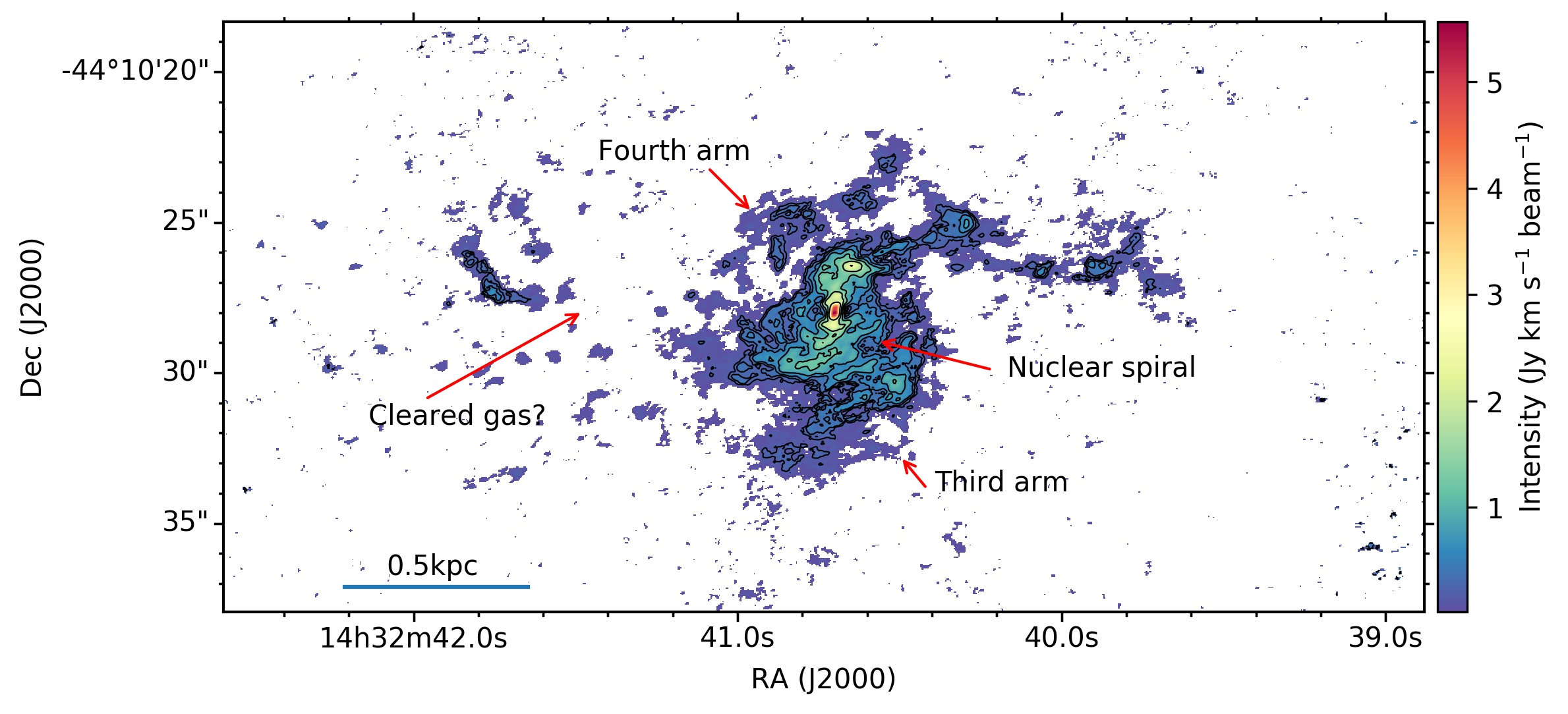}
\caption{Top panel: Optical image of NGC~5643 from the Carnegie-Irvine
  Galaxy Survey \citep{Ho2011}. The red
    rectangle marks the approximate size of the ALMA map below. Lower
    panel: ALMA CO(2--1) natural-weighted integrated intensity map
  of  NGC~5643  produced with a $3\sigma$ clipping and shown in a linear
  scale to emphasize the large scale emission. The contours are shown in a logarithmic scale with the first
  contour at $0.2\,{\rm Jy\, km\,s}^{-1}\,{\rm beam}^{-1}$ and the
  last contour at $2.5\,{\rm Jy\, km\,s}^{-1}\,{\rm beam}^{-1}$. The
  FoV of the image is 
  approximately $39.9\arcsec \times 19.9\arcsec$ ($\sim 3.3\,{\rm kpc}\times 1.6\,{\rm kpc}$)  and the angular
  resolution $0.26\arcsec \times 0.17\arcsec$ at 
  ${\rm PA_{\rm beam}}=-69.6\arcdeg$. We applied the primary beam (27\,\arcsec) correction
  to the data.}\label{fig:ALMAfullFoV}
\end{figure*}

\section{Observations} \label{sec:obs}

\subsection{ALMA Band 6 observations}\label{sec:ALMAobs}
We obtained Band 6 ALMA observations of NGC~5643 on 24 December 2016 and 18 July 2017 using the 12-m array in compact (baselines between 15 and 492\,m) and extended (baselines between 17 and 3700\,m) configurations through the project 2016.1.00254.S (PI: A. Alonso-Herrero). The on-source integrations times were 11 and 36\,min, respectively. 
We defined two spectral windows of 1.875\,GHz bandwidth
(3.9\,MHz~$\sim$ 5\,km\slash s channels), namely one at
$\sim$228\,GHz at the observed frequency of the $^{12}$CO(2--1) transition  and the
other at observed frequency of $\sim$230\,GHz to measure the sub-millimeter continuum. 

We calibrated the data using the ALMA reduction software CASA (v.4.7; \citealt{McMullin2007}). For both configurations, we used J1427-4206 as bandpass and phase calibrator. The amplitude was calibrated using J1427-4206 for the extended configuration assuming a flux density of 1.92\,Jy at 228.6\,GHz and a spectral index of --0.59. For the compact configuration, we used Callisto  as amplitude calibrator assuming the Butler-JPL-Horizons 2012 model.

For the CO(2--1) spectral window, we subtracted the continuum (rest-frame 232\,GHz) in the
$uv$ plane by fitting the continuum with a constant in the line free
channels. Then, we combined and cleaned the data from the two
configurations using the CASA {\it clean} task. The output frequency reference frame was set to the kinematic local standard of rest (LSRK).

We produced three sets of cleaned data 
For the first one we used natural weight to emphasize the large scale emission of  NGC~5643.
For the other two we used the Briggs weighting
\citep{Briggs1995PhDT}, with robustness parameters of $b=0.8$ 
and $b=-0.5$ to increase sequentially the angular resolution
at slightly decreased  sensitivity. For the $b=-0.5$ data set we kept
the original spacings of the frequency channel whereas for
the $b=0.8$ set we rebinned the data by a factor of 3 in frequency to increase the
signal-to-noise ratio. This resulted in an approximate resolution in velocities of
5 \,km\,s$^{-1}$ and  CO(2--1)  full-width half-maximum (FWHM) of 
0\farcs16$\times$0\farcs11 with beam position angle  (${\rm PA}_{\rm beam}$) 
of --67\arcdeg\ for the $b=-0.5$ data set, and 
15\,km\,s$^{-1}$ and 
FWHM
of 0\farcs26$\times$0\farcs17 with ${\rm PA}_{\rm beam}=-58$\arcdeg\ for the $b=0.8$ data set.  For the continuum images, the beams 
FWHM are 0\farcs16$\times$0\farcs10 at ${\rm PA}_{\rm beam}
=-65\arcdeg$ for $b=-0.5$ and 
and 0\farcs23$\times$0\farcs13 at ${\rm PA}_{\rm beam}=-68\arcdeg$ for
$b=0.8$.
The pixel sizes were set to 0\farcs03 and 0\farcs04 to properly sample the
   beam sizes, for the $b=-0.5$ and $b=0.8$ data sets, respectively.
The achieved 1$\sigma$ sensitivities are $\sim$1.2 and
$\sim$0.5\,mJy\,beam$^{-1}$\,channel$^{-1}$ in the CO(2--1) cubes and
$\sim$76 and $\sim$37\,$\mu$Jy\,beam$^{-1}$ in the continuum
images for the $b=-0.5$ and $b=0.8$ data sets respectively.  For all three data cubes
we produced maps of  the  CO(2--1) integrated intensity, mean
velocity field and velocity dispersion. For all these maps we used pixels at all frequencies with
detections $>3\sigma$. We applied the primary beam (FWHM=27\arcsec) correction to the
data.

In Figure~\ref{fig:ALMAfullFoV} (lower panel)
we show the  integrated CO(2--1) molecular line  map created with the natural-weight data.
The field of view (FoV) is approximately $39.9\arcsec \times
19.9\arcsec$ ($\sim 3.3\,{\rm kpc}\times 1.6\,{\rm kpc}$)  with an angular
resolution of $0.26\arcsec \times 0.17\arcsec$ at ${\rm PA}_{\rm
  beam}=-69.6\arcdeg$. We note that the primary beam of the
observations  is smaller than this FoV. However, most of the detected
emission is well inside the FoV of the primary beam.  The line map
shows a bright nuclear source, a nuclear two-arm spiral already seen in optical to
near-IR color maps as a dusty nuclear spiral extending for
  several arcseconds \citep{Quillen1999,
  Martini2003, Davies2014} as well as
emission in the
spiral arms on larger scales. We measure a CO(2--1) line intensity over a region
$40\arcsec \times 40\arcsec$ of 480\, Jy km s$^{-1}$. For comparison 
the single-dish measurement for a beam size of $30\arcsec$ is $912
\pm 60\,$ Jy km s$^{-1}$ \citep {Monje2011}. Thus we recovered
approximately 50\% of the single-dish observation flux which is typical of this
kind of comparisons.

\subsection{Archival VLT/MUSE integral field spectroscopy}\label{sec:MUSEobs}
NGC~5643 was observed with the Multi Unit Spectroscopic
Explorer \citep[MUSE, ][]{Bacon2010} on the Very Large Telescope (VLT)  of the
European Southern Observatory (ESO) on 12 May 2015. These observations were part of the program
095.B-0532(A) (PI: Carollo). 
  We downloaded the pipeline processed data cube from the ESO data archive. We estimate a
  seeing of $\sim$0\farcs5 from the foreground stars in the FoV. Four
  dithered exposures of 875\,s were taken and later combined into a
  single cube by the pipeline. This final combined cube covers a
  1\arcmin$\times$1\arcmin\ FoV with a pixel size of 0\farcs2. The
  observed spectral range is 475--935\,nm and the spectral resolution
  varies between 
$R=\lambda/\Delta\lambda_{FWHM}\sim 1770$  at 480\,nm and $R=3590$ at 930\,nm.

  We produced line emission maps and velocity fields of the [O\,{\sc iii}]$\lambda$4959 and $\lambda$5007 doublet
  (the latter will be refered to as [O\,{\sc iii}] hence forward),  the [N\,{\sc ii}]$\lambda$6549 and
  $\lambda$6583 doublet, and
 H$\alpha$ at 6563\,\AA. We fitted a single Gaussian profile to each of these transitions in each spaxel. The underlying local continuum was approximated by a straight line. The flux ratio of the doublets was
  fixed to the expected theoretical ratio of 0.35 and 0.34 for the
  [O\,{\sc iii}] and [N\,{\sc ii}]  doublets respectively. The [N\,{\sc ii}] and H$\alpha$ transitions were fitted simultaneously.

\begin{figure*}
%\epsscale{1.05}
%\hspace{1cm}
%\plottwo{figure2a.eps}{figure2b.eps}
%\vspace{-1.1cm} 
\includegraphics[width=0.5\textwidth]{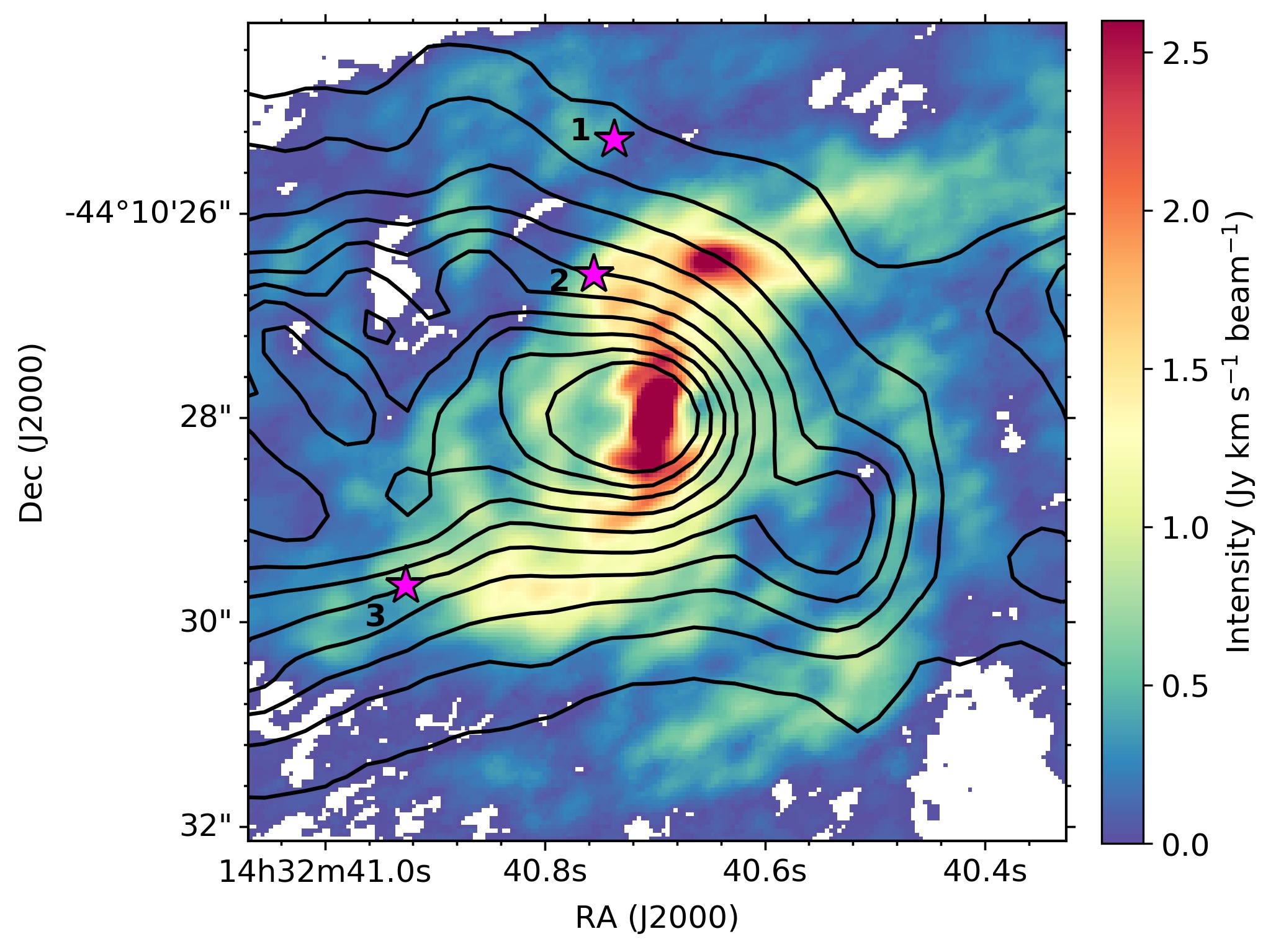}
%\hspace{-1.2cm}
\includegraphics[width=0.5\textwidth]{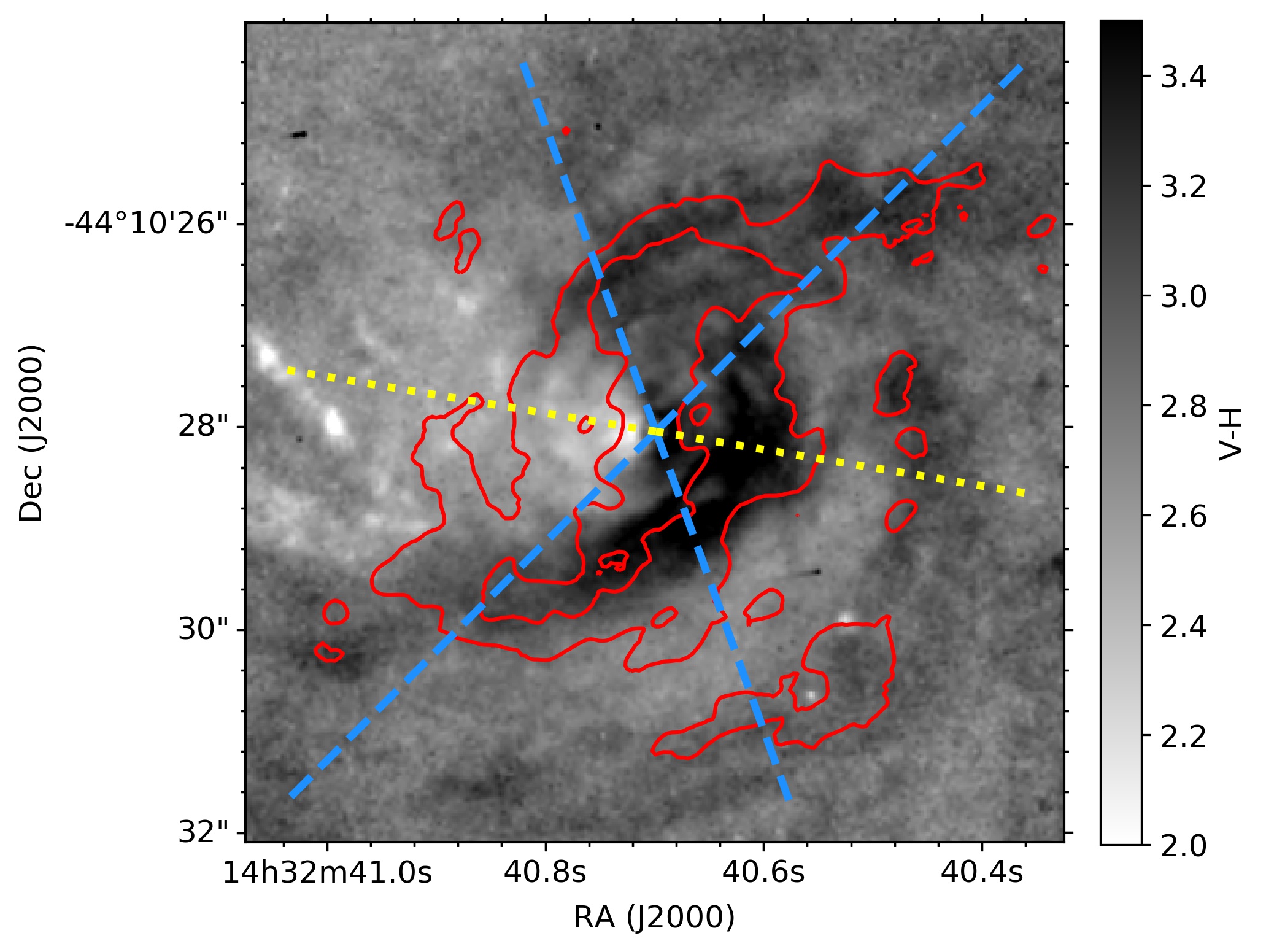}
%\vspace{-1.5cm}
 \caption{Left panel. In color is the ALMA CO(2--1)  integrated intensity map of
  NGC~5643  in a linear
  scale produced from the $b=0.8$ data cube  (angular
resolution of $0.26\arcsec \times 0.17\arcsec$  at ${\rm PA}_{\rm
  beam}=-58.9\arcdeg$) with a $3\sigma$ clipping. We show a FoV of $8\arcsec \times 8\arcsec$  (central $655\,{\rm pc} \times 655\,{\rm
  pc}$) for easy comparison with the H$_2$ $2.12\,\mu$m
observations of \cite{Davies2014}.   
The black contours are the MUSE [O\,{\sc iii}]
continuum-subtracted line emission. The star-like symbols
  mark the location of the regions in Figure~\ref{fig:lineprofiles}.
Right panel. In grey scale is the $V-H$ color map produced
with {\it HST} images \citep{Martini2003, Davies2014} where dark colors
indicate regions with higher extinction. We show the ALMA CO(2--1)
  emission with two red contours  to guide the eye. We also mark
    with the blue dashed lines the approximate boundary of the
    ionization bicone (semi-opening angle of 55\arcdeg) and with the
    yellow dotted line the bicone axis (the cone inclination out of
    the plane of the sky towards Earth is  $\sim
    25\arcdeg$,  and $\sim
    40\arcdeg$ \, with respect to the galaxy disk plane), as modelled by 
\cite{Fischer2013}.}\label{fig:ALMAOIIIcolormap}
\end{figure*}

Finally we aligned the MUSE images using the peak of the continuum
images to match  the coordinates of the ALMA continuum peak at rest-frame
232\,GHz (1.3\,mm) 
RA(J2000) =14$^{\rm h}$ 32$^{\rm m}$ 40.70$^{\rm s}$ and Dec(J2000)
= $-44\arcdeg$
10$\arcmin$ 27.9$\arcsec$.
\section{Circumnuclear region}\label{sec:circumnuclear}

In this section we focus on the CO(2--1) emission in the circumnuclear
region of NGC~5643 over a FoV of 
$8\arcsec \times 8\arcsec$  (central $655\,{\rm pc} \times 655\,{\rm
  pc}$) and use the $b=0.8$ data set. We selected this region to
compare the cold molecular gas emission with the VLT/SINFONI integral field spectroscopy study of this
galaxy done by \cite{Davies2014} with this FoV. Nevertheless, as can
be seen from Figure~\ref{fig:ALMAfullFoV} (lower panel), 
most of the ALMA CO(2--1) emission ($\sim 80\%$)
arises from this region. Based on the kinematics of the hot molecular gas
line H$_2$ at $2.12\,\mu$m \cite{Davies2014} found
evidence of outflowing material on scales 
$1-2\,\arcsec$ from the nucleus. This material appears to be at the edge of the
eastern ionization cone probed by the [O\,{\sc iii}] emission
observed with the {\it Hubble Space Telescope} ({\it HST)} by \cite{Simpson1997} and more recently
with MUSE by \cite{Cresci2015}.

\subsection{Morphology}\label{sec:morph8x8}

Figure~\ref{fig:ALMAOIIIcolormap}  (left panel) shows in color  the
CO(2--1) velocity-integrated intensity map.
The brightest emission comes from the central region
 (see Section~\ref{sec:morphnuclear} for a detailed discussion) from which
the nuclear two-arm spiral structure is
clearly connected. The spiral structure extends for more than 10\arcsec \,
on both sides of the galaxy (see lower panel of Figure~\ref{fig:ALMAfullFoV}),
although the east side spiral arm shows a region  void of CO(2--1)
emission gas. We will come back to this in Section~\ref{sec:conclusions}. 
The spiral arms are oriented in an almost east-west direction as is
  the large-scale stellar bar \citep[see
  e.g.][and also upper panel of Figure~\ref{fig:ALMAfullFoV}]{Mulchaey1997, Jungwiert1997}.
There is also CO(2--1) emission in  a  third
 spiral-like structure in the circumnuclear region
 located to the west of the nucleus. We marked this third arm
 in Figure~\ref{fig:ALMAfullFoV} (lower panel). This structure was 
first detected in H$_2$ at $2.12\,\mu$m emission by \cite{Davies2014}.
They interpreted  as a transient structure resulting from a recent perturbation to
the galaxy. Alternatively it might be attributed to the far side of the ionization
cone (that is, the counter-cone) where it is bisecting the galaxy and perturbing the ambient
gas. In the nuclear region the tips of the nuclear spiral arms connect with
  dusty arc-like structures $\sim 1.4\arcsec$ in size. In the innermost region a disk-like
  structure ($\sim 0.6\arcsec$ diameter) in the north-south direction is detected. These will
be discussed in detail in Section~\ref{sec:nuclear}.

\begin{figure*}
\epsscale{1.15}
\plotone{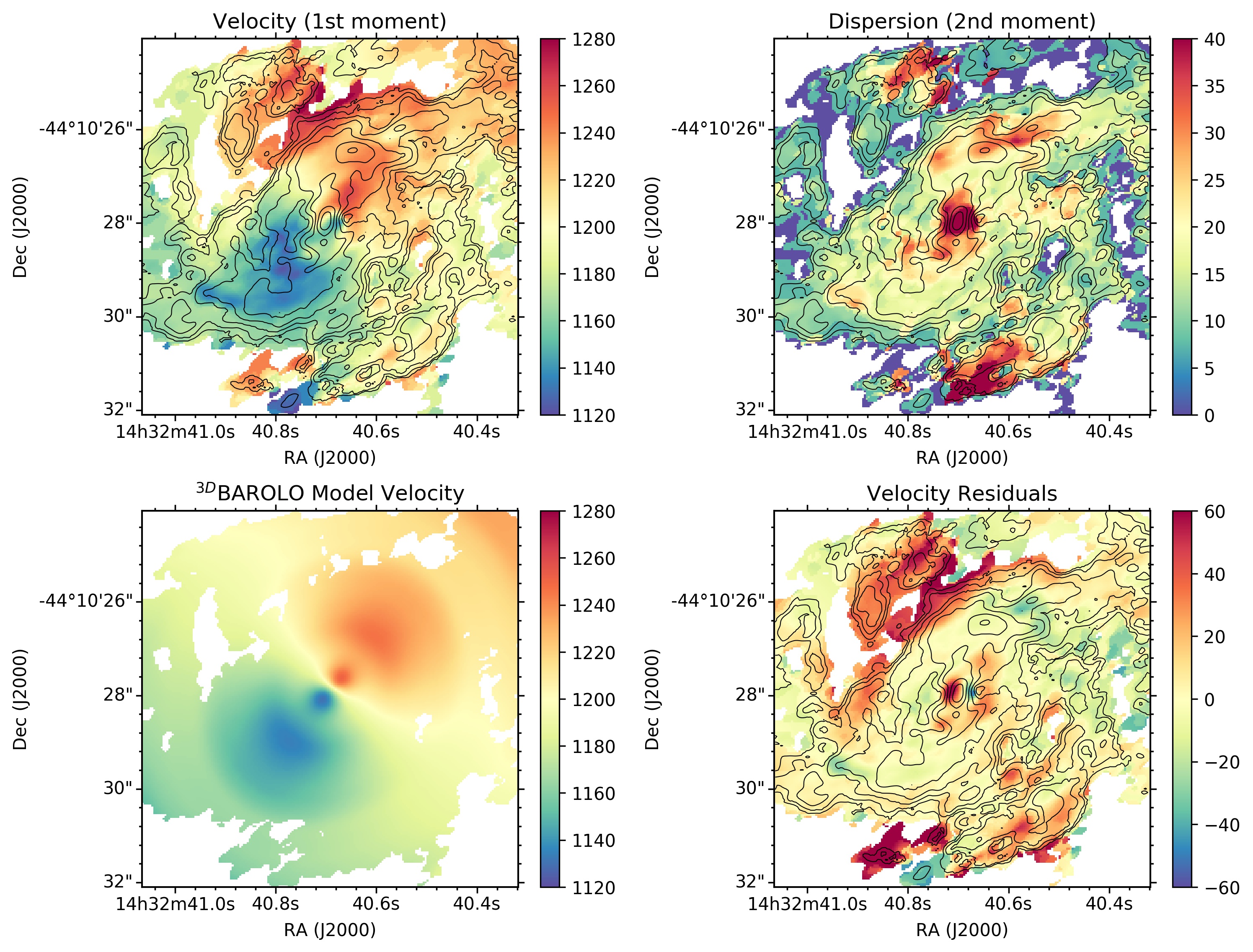}
\caption{Modeling of the ALMA CO(2--1) kinematics with  $^{\rm
    3D}$BAROLO using the $b=0.8$ data cube (angular
resolution of $0.26\arcsec \times 0.17\arcsec$  at ${\rm PA}_{\rm
  beam}=-58.9\arcdeg$) and a FoV of $8\arcsec \times 8\arcsec$
(as in Figure~\ref{fig:ALMAOIIIcolormap}). The top left and right panels are the mean
  velocity field (1st moment map) and velocity dispersion field (2nd
  moment map), respectively. The bottom left
  panel is the disk model and the bottom right panel is the residual
  mean-velocity field map.  The vertical color bars are velocities or
  velocity dispersions in km s$^{-1}$. The black contours are the 
CO(2--1) line intensity in a linear scale. }\label{fig:BAROLO8x8}
\end{figure*}

We superimposed on the CO(2--1) map in Figure~\ref{fig:ALMAOIIIcolormap}  as black contours our MUSE
continuum-subtracted [O\,{\sc iii}] line emission map (see Section~\ref{sec:MUSEobs}).  We note that
this new MUSE data set was observed under better seeing
conditions  than the \cite{Cresci2015} one (0.5\arcsec\, versus 0.88\arcsec).
As discussed by these
authors and also seen in our figure, the ionization cone traced by the  [O\,{\sc iii}]
emission is more clearly seen to the east of the nucleus due to
obscuration produced by the host galaxy in the west
direction. However, the MUSE  [O\,{\sc iii}] image also traces the counter-cone
emission on scales of a few arcsec to the west of the nucleus \citep{Cresci2015}.
%For reference we
%also plotted as blue contours  the [O\,{\sc iii}] map produced
%with the {\it HST}/WFPC2 F502N narrow-band filter after subtracting the
%F547M medium-band filter image \citep[see][for full
%details]{Simpson1997}. While the {\it HST} image has better angular
%resolution than the MUSE image it misses a large fraction of the
%diffuse [O\,{\sc iii}] emission and more importantly the redshifted emission
%that falls outside of the narrow-band F502N filter.
This comparison
shows that most of the CO(2--1) bright emission seems to {\it avoid} the regions of
ionized [O\,{\sc iii}] gas. This would suggest that the CO(2--1) emission is in the plane of
the galaxy whereas most of the [O\,{\sc iii}] emission is outflowing
perpendicular  to the accretion disk and intercepts the host
  galaxy disk, since the ionization cone has an inclination of
  approximately 40\arcdeg \, with respect to the galaxy disk \citep[see also][and
Section~\ref{sec:kinem8x8}]{Fischer2013}. Similar situations of the molecular
gas avoiding the ionization cone have been reported in the literature
for NGC~4253 and M~51
\citep[see ][respectively]{Krause2007, Querejeta2016}.

The right panel of Figure~\ref{fig:ALMAOIIIcolormap}  is the {\it HST}
$V-H$ color map (in grey scale) from \cite{Martini2003} and
\cite{Davies2014}.  We overlaid on this image  two contours showing the CO(2--1) integrated
emission (that of the left
panel) to guide the eye.  Clearly  the nuclear spiral
structure outlined by the nearly horizontal dust lanes  (dark $V-H$
regions) entering the nuclear region is coincident with the CO(2--1) molecular gas emission. 
The third spiral-like structure is traced by  both the CO(2--1) cold gas
emission and the  dust lanes.  Also there is evidence that this third spiral structure might
have a fainter counterpart (i.e., a {\it fourth} arm) on the east side of the galaxy again seen
in both dust and cold molecular gas (see also
Figure~\ref{fig:ALMAfullFoV} for a larger FoV).

\subsection{Kinematics}\label{sec:kinem8x8}
We used the $^{\rm 3D}$BAROLO code \citep{DiTeodoro2015}  to model
the CO(2--1) kinematics of the central $8\arcsec \times 8\arcsec$
region of
NGC~5643 using the ALMA $b=0.8$ data cube. $^{\rm 3D}$BAROLO was
designed to fit simple disk models using 3D tilted rings to a variety
of emission line data, including ALMA data cubes. The fit is done in
two steps. During the first  step for
each ring $^{\rm 3D}$BAROLO fits the free parameters, namely, 
the kinematic center, the systemic velocity, the disk
inclination and PA of the major axis,  the scale height of the disk, the
circular velocity and velocity
dispersion. The second step fixes the kinematic center and systemic
velocity to the mean fitted values from the first iteration and finds the best
solution for the other free parameters.  $^{\rm 3D}$BAROLO also allows
the user to fix any of the above parameters during the first
iteration. After fitting the data cube, $^{\rm 3D}$BAROLO creates,
among other products, a map of the observed 
velocity-integrated intensity (0th moment), a map of the observed mean velocity field
(1st moment) and velocity dispersion
field (2nd moment) as well as the same maps for the best fit model.

We first run $^{\rm 3D}$BAROLO on the CO(2--1) data cube 
using the disk PA and inclination
derived from the near-infrared stellar kinematics by \cite{Davies2014}
over the same FoV allowing for
relatively small variations (approximately $\pm 20\arcdeg$ for the
disk PA and $\pm 10\arcdeg$ for the disk inclination). After the first
run we fixed the disk PA and inclination to the average fitted values
of ${\rm PA}_{\rm disk}= 320\arcdeg$ and $i=35\arcdeg$, which are  compatible with
those inferred from the stellar kinematics \citep{Davies2014}. We also fixed the systemic
velocity to $v_{\rm sys}=1194\,{\rm
  km\,s}^{-1}$ (mean value of the first run)
and rerun $^{\rm 3D}$BAROLO to fit the observed velocity field.

\begin{figure}
\epsscale{1.2}
\plotone{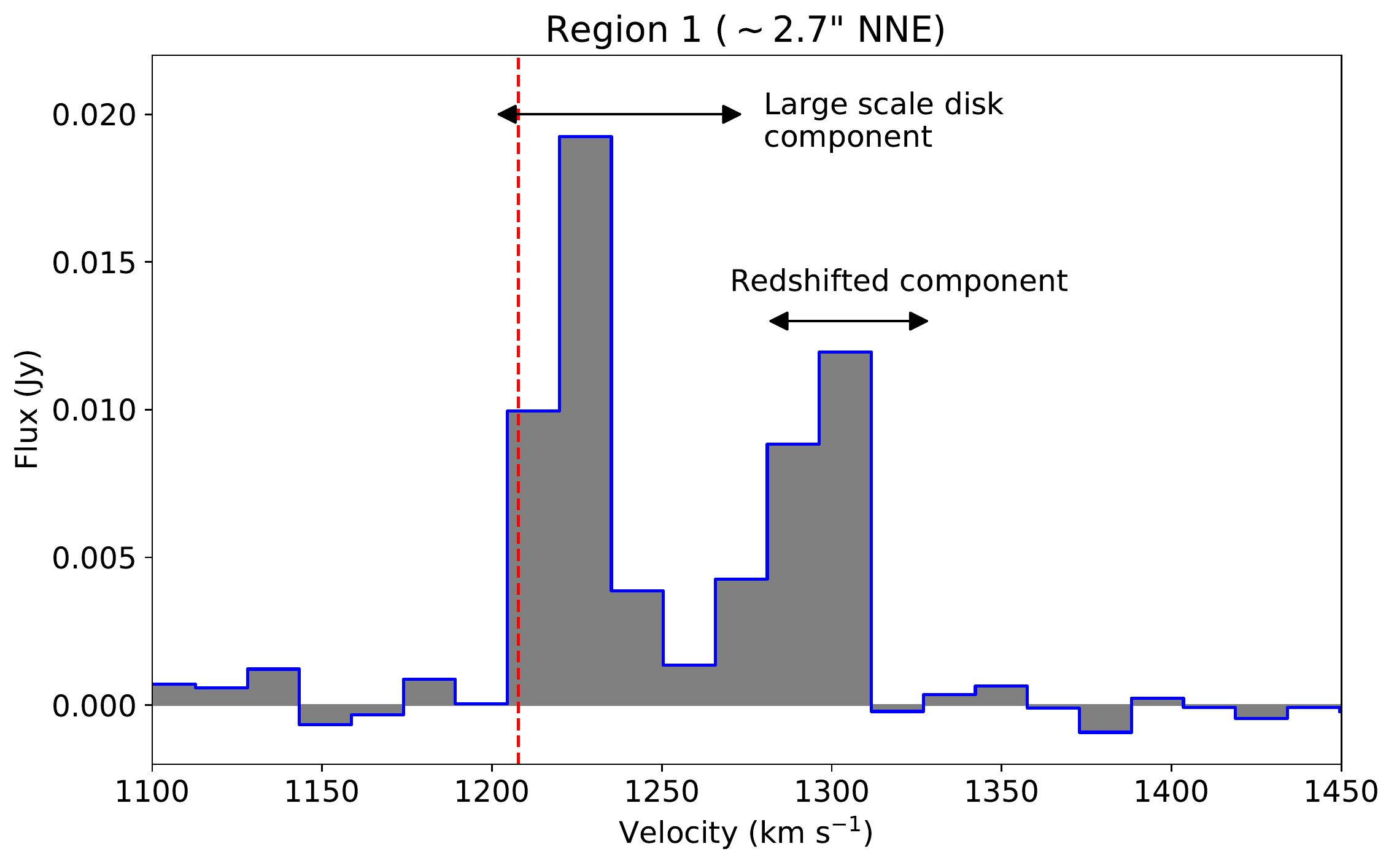}
\plotone{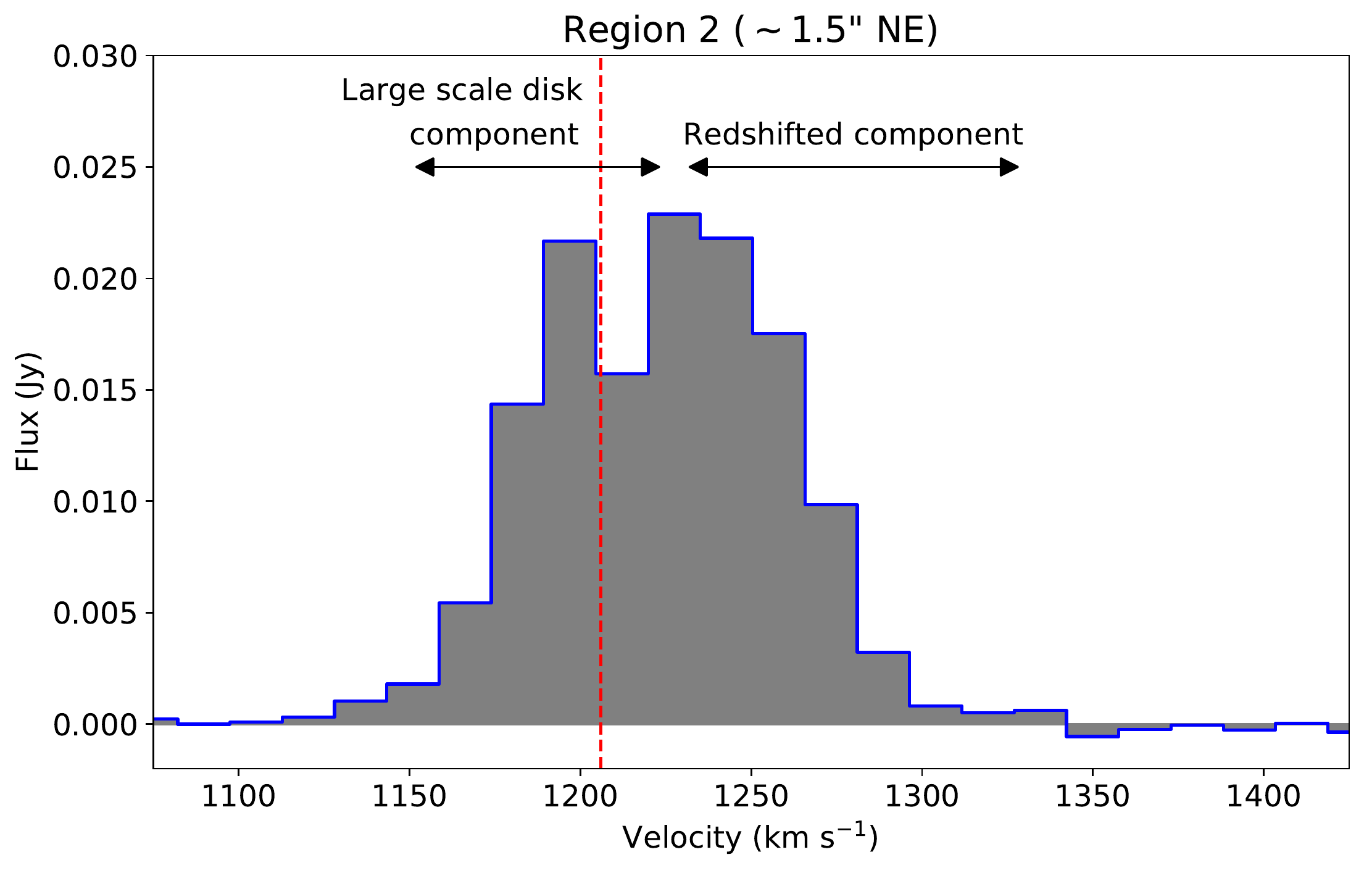}
\plotone{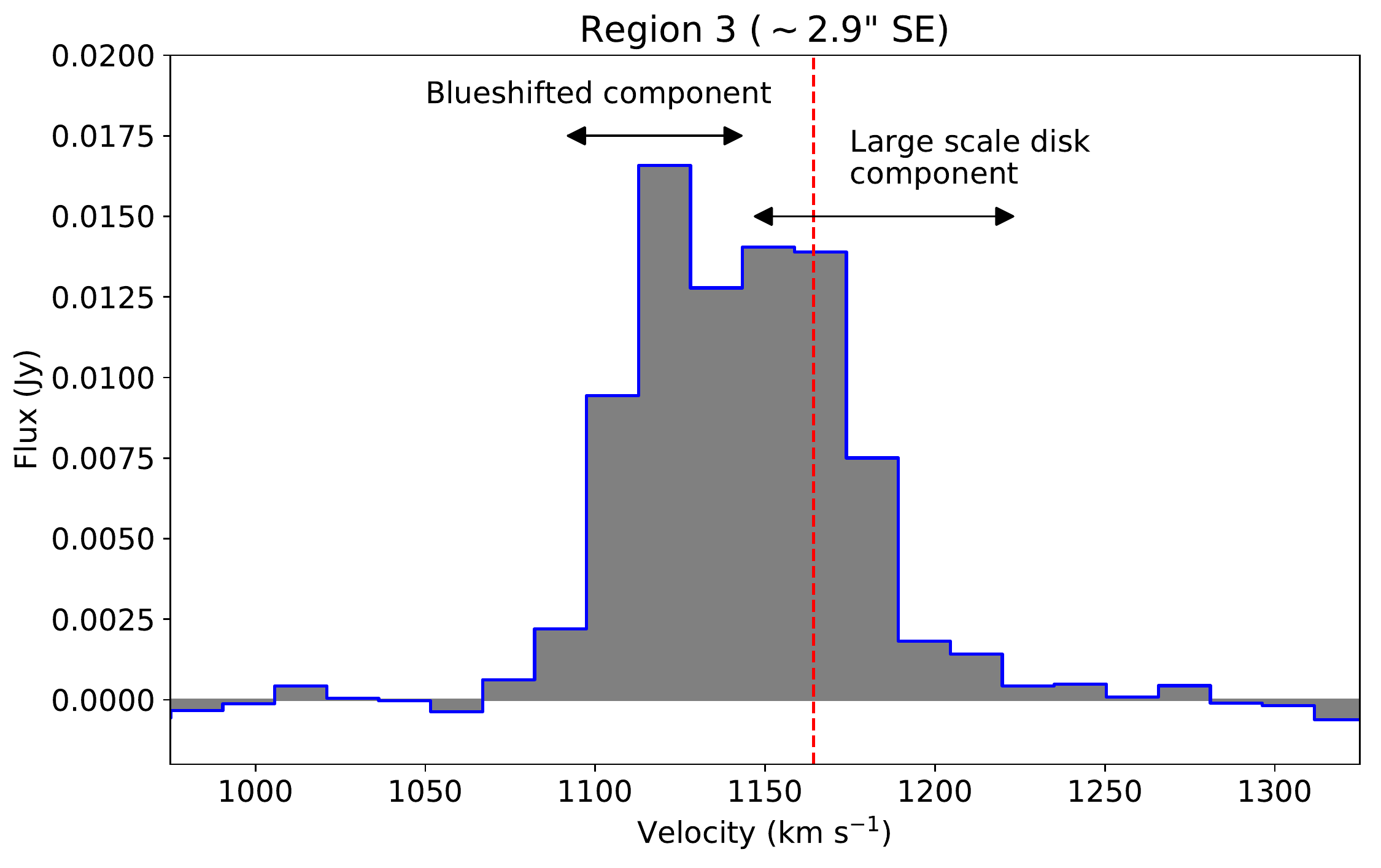}
\caption{CO(2--1) line profiles extracted from
  the $b=0.8$ data cube at three different locations with
  large velocity residuals and/or large velocity dispersion. The velocity resolution
  of the data is approximately $15\,{\rm km\,s}^{-1}$. The vertical dashed
  lines are the expected disk circular velocities from
the $^{\rm 3D}$BAROLO fit. 
The coordinates (J2000) of the regions are as follows.
Region 1:  RA=14$^{\rm h}$ 32$^{\rm m}$ 40.73$^{\rm s}$ and Dec =
$-44\arcdeg$ 10$\arcmin$ 25.2$\arcsec$, 
Region 2:  RA= 14$^{\rm h}$ 32$^{\rm m}$ 40.75$^{\rm s}$ and Dec =
$-44\arcdeg$ 10$\arcmin$ 26.6$\arcsec$, and 
Region 3:  RA= 14$^{\rm h}$ 32$^{\rm m}$ 40.93$^{\rm s}$ and Dec =
$-44\arcdeg$ 10$\arcmin$ 29.6$\arcsec$. }\label{fig:lineprofiles}
\end{figure}

The results are shown in Figure~\ref{fig:BAROLO8x8}. The top panels
are the  $^{\rm 3D}$BAROLO maps of the observed mean CO(2--1)
velocity field (1st moment map, left) and
velocity dispersion (2nd moment map, right). These can be compared directly with the
H$_2$ $2.12\,\mu$m hot molecular gas maps in Figure~8 of
\cite{Davies2014}.  Both the mean velocity fields of the cold and hot
molecular gas are 
similar and display a clear rotational pattern with evidence of some
non-circular motions.  The velocity dispersion maps of the cold and
hot  molecular gas show a
peak at the nucleus position with a value of  $\sigma({\rm CO(2-1))} \sim 
60\,{\rm km\,s}^{-1}$. However, the  H$_2$ map shows the highest
velocity dispersion value, $\sigma({\rm H}_2) \sim 100\,{\rm km\,s}^{-1}$, 
 at about 2$\arcsec$ to the northeast of the
nucleus. This peak is not as clearly seen in the map of CO(2--1) velocity
dispersion due to low signal-to-noise ratio in that region of our map.
Nevertheless, there is a region to the north-northeast of the AGN
at about $3\arcsec \sim 246\,$pc with
a CO(2--1) velocity dispersion value similar to that of the nucleus (see below).

We constructed a CO(2--1) residual mean-velocity field map  (see
Figure~\ref{fig:BAROLO8x8}, bottom right panel)
by subtracting the  $^{\rm 3D}$BAROLO disk model (bottom left panel of
Figure~\ref{fig:BAROLO8x8}) from the $^{\rm 3D}$BAROLO first moment
map. Some of the residual velocity
  field seen in the central few
  arcseconds could be an artifact of the simple disk geometry
  assumed in our modelling with $^{\rm 3D}$BAROLO. In particular, a model
  which accounted for the presence of a bar potential would predict better
  the streaming motions linked to the gas response to the bar along
  the leading edges of the nuclear spiral \citep[see for instance, ][]{Emsellem2001}.
  However, it would not be able to reproduce the strong redshifted velocity residuals
  ($\sim 40-60\,{\rm km\,s}^{-1}$) seen $\sim 2-3\arcsec$ northeast of the AGN.
  This redshifted component
  was also observed in the velocity
  field of the 
  hot molecular gas H$_2$ at $2.12\,\mu$m \citep{Davies2014}
and shows high CO(2--1)  velocity dispersion.

To look further into the gas kinematics of  regions with the large velocity residuals
with respect to a simple rotating disk
and/or velocity dispersions, we extracted
CO(2--1) line profiles at  three positions. The first two have
redshifted components and are located approximately $2.7\arcsec$
northnortheast 
and   $1.5\arcsec$ northeast ($\sim 221$ and $123\,$pc respectively,
  and see Figure~\ref{fig:ALMAOIIIcolormap}
for the locations) from the peak
of the CO(2--1) emission. The top and middle panels of
Figure~\ref{fig:lineprofiles} show the line profiles obtained with a square 
aperture of 10 pixels or 0.4\arcsec.  Both
regions clearly show two distinct velocity components, one associated with the rotation of
the large scale disk and a second  one redshifted by approximately $30-45\,{\rm km\,s}^{-1}$.
The morphology of the nuclear spiral  structure indicates that the rotation  in the disk of
the galaxy is  counterclockwise. Assuming that the non-circular motions seen in 
these regions to the northeast of the center are co-planar with the galaxy
disk then the redshifted components imply  radial outward movements in the disk of the galaxy
near the edge of the [O\,{\sc iii}] ionization cone
(see Figures~\ref{fig:ALMAOIIIcolormap} and \ref{fig:BAROLO8x8}).
The simplest
explanation for the CO(2--1) velocity residuals would be radial
movements of material that is
being pushed outwards in the galaxy disk by the
outflowing material traced by the [O\,{\sc iii}] emission.

The  third region is in the nuclear
spiral at  $\sim 2.9\,$\arcsec \ southeast of the nucleus
(see Figure~\ref{fig:ALMAOIIIcolormap})
and presents blue velocity residuals and an excess velocity
dispersion, which are explained by the presence of two velocity components
(see bottom panel of Figure~\ref{fig:lineprofiles}).
This
location is in a region where the [O\,{\sc iii}] outflow (identified by the presence
  of blueshifted components)
  and the radio jet appear to be
interacting with the disk of the galaxy \citep{Cresci2015}. We note that this CO(2--1)
blueshifted region is not coincident with the H$\alpha$  knots resulting
from positive feedback induced by the jet that are located
further away from the nucleus.

\begin{figure*}
%\hspace{0.5cm}
\includegraphics[width=0.5\textwidth]{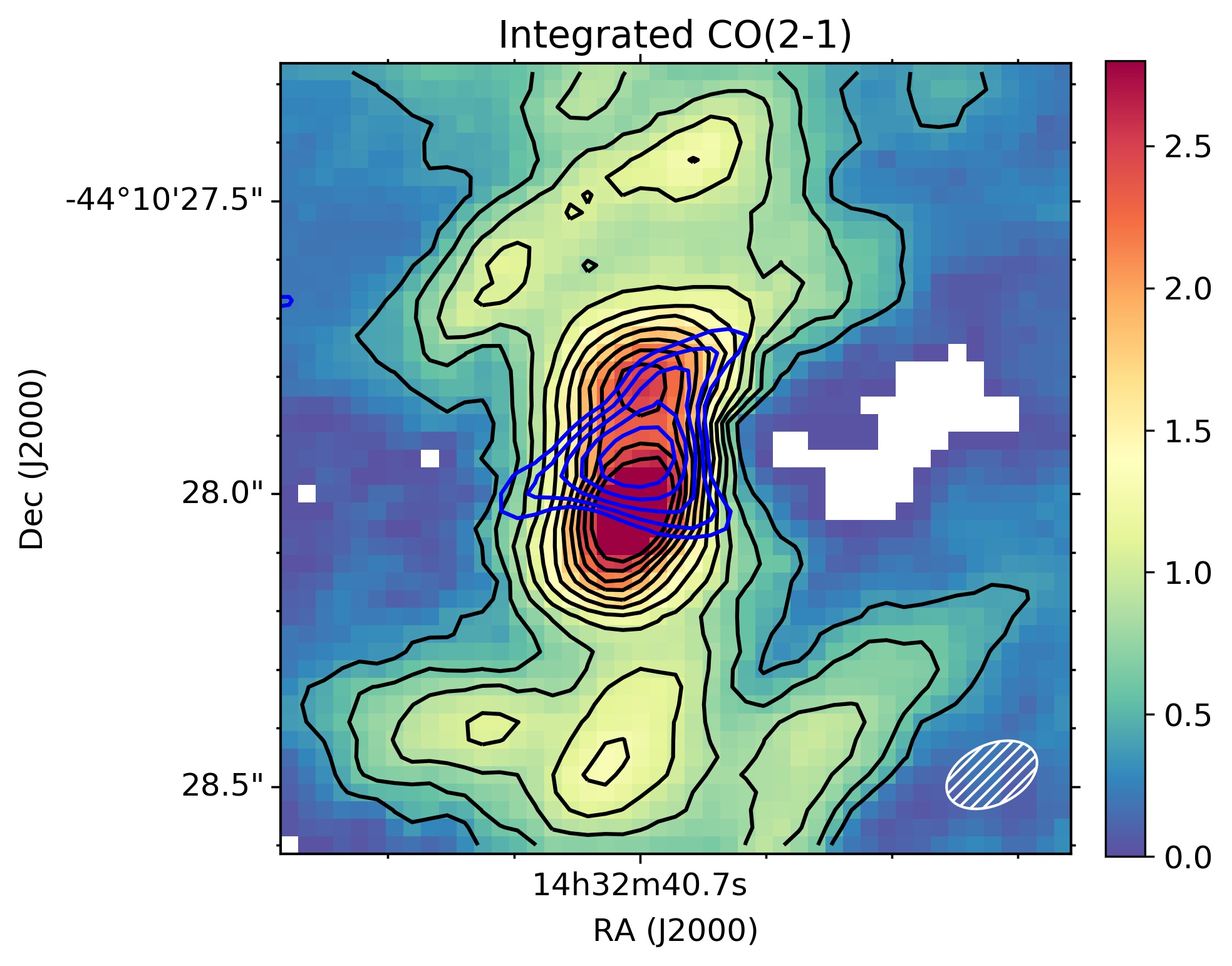}
\hspace{-0.05cm}
\includegraphics[width=0.5\textwidth]{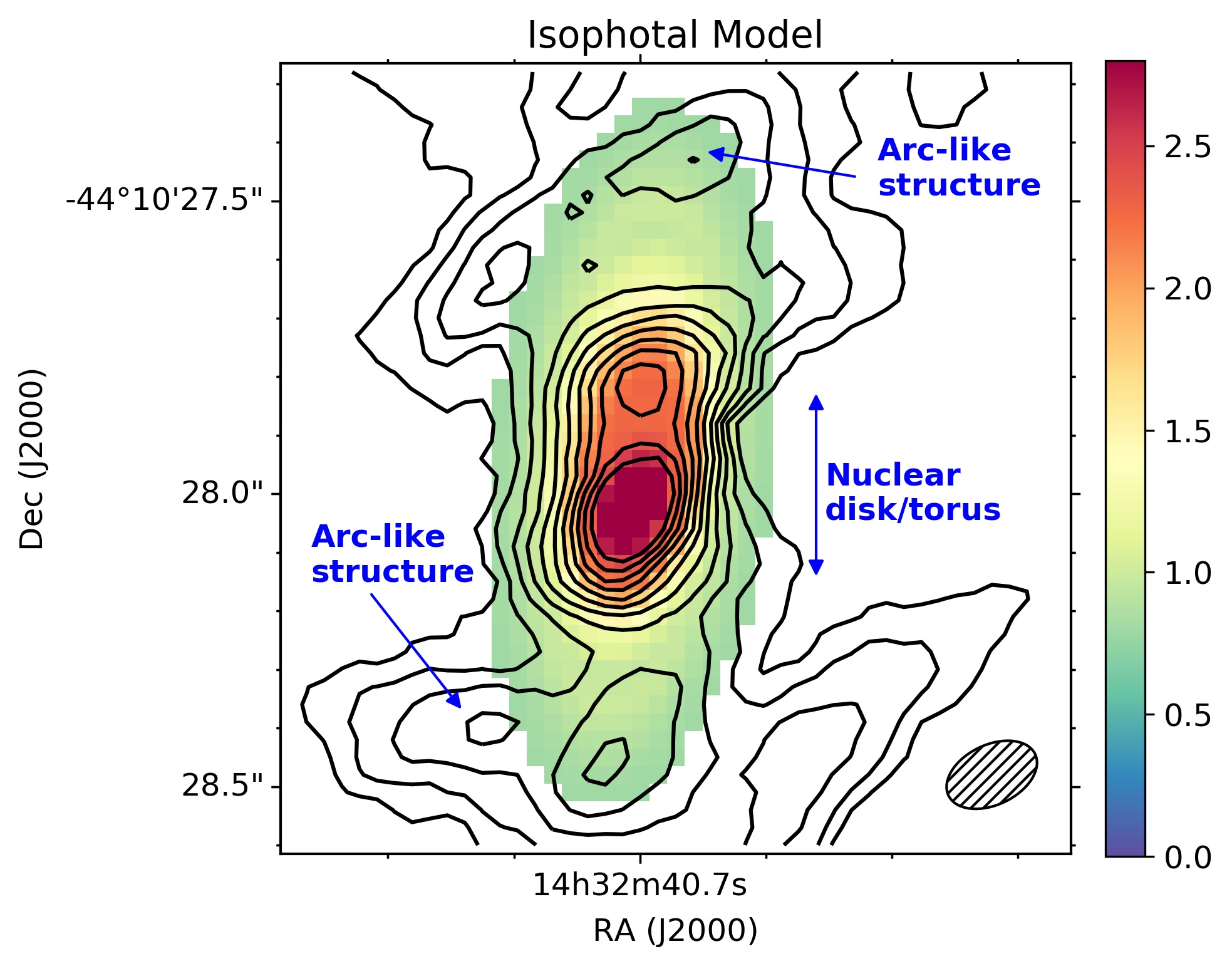}
%\vspace{-1.5cm}
\caption{The left panel shows in color in a
  linear scale  the observed ALMA CO(2--1)
  $b=-0.5$ integrated intensity map of the nuclear region of 
  NGC~5643. The black contours are also the CO(2--1) intensity in a
  linear scale with the lowest contour  at 0.6\,Jy km s$^{-1}$
  beam$^{-1}$ and 
  contour steps of 0.2\,Jy km s$^{-1}$ beam$^{-1}$. The blue contours are the Band 6 rest-frame 232\,GHz
  continuum. The right panel shows in color the isophotal model fitted to the nuclear
  disk CO(2--1) emission. The black contours are as in the left panel. We mark the
    location of the nuclear arc-like structures and the disk/torus. For the latter the size of the
    arrow indicates the measured FWHM in the north-south direction not corrected for the ALMA beam (ellipse, 
  0\farcs16$\times$0\farcs11 at
  $PA_{\rm beam}=-67$\arcdeg). }\label{fig:ALMAsmallFoV} 
\end{figure*}

\section{The nuclear obscuring disk}\label{sec:nuclear}
To study the properties of the nuclear emission we use the ALMA
$b=-0.5$ data
cube with the highest angular resolution of $0.16\arcsec
\times 0.11\arcsec$ which corresponds to $13\,{\rm pc } \times 9\,{\rm
pc}$ at the assumed distance of NGC~5643. As already seen from 
Figures~\ref{fig:ALMAfullFoV} and \ref{fig:ALMAOIIIcolormap},  NGC~5643 shows bright  CO(2--1) line emission arising
from the nuclear regions.

\subsection{Morphology}\label{sec:morphnuclear}
We focus here on the central $1.35\arcsec
\times 1.35\arcsec$ region which corresponds approximately to the
central $110\,{\rm pc} \times 110\,{\rm pc}$, and
encloses the arc-like structures and the nuclear disk (see the map of the CO(2--1) integrated
intensity in Figure~\ref{fig:ALMAsmallFoV}). The 
arc-like structures trace where the nearly horizontal nuclear spiral
arms (oriented as the large-scale bar) come in
(see also Figure~\ref{fig:ALMAOIIIcolormap}). These are coincident with dusty features suggesting the
direction of the  inflowing material \citep{Davies2014}. The presence
of these straight dust
lane/gas features near the nucleus  are predicted by hydrodynamical models
of gas flow in a barred potential \citep{Regan1999,
  Maciejewski2004}.

The innermost region shows a disk-like
morphology.  This nuclear disk is fully resolved with a measured FWHM of
$0.32\arcsec$  in the approximate north-south direction and
$0.24\arcsec$ in the east-west direction
which corresponds to a projected size of $26\,{\rm pc}  \times
20\,{\rm pc}$. These values  have not been corrected for the 
beam size of the observations.  It also appears to be tilted with
  respect to the large scale galaxy disk (see Section~\ref{sec:kinemnuclear}).
The 
presence of this nuclear disk was previously suggested by the dusty structure in the nuclear
region \citep{Davies2014, Cresci2015}. 
Alternatively we could interpret the CO(2--1) nuclear structure as an
inner bar (see discussion in Section~5.2).
%However, \cite{Mulchaey1997} detected in the near-infrared $K$-band a large scale bar in this galaxy
%at $PA=85\arcdeg$ with a length of $67\arcsec$ (see also upper panel
%of Figure~\ref{fig:ALMAfullFoV}) but found no evidence of the
%presence of a nuclear bar.

The nuclear disk of NGC~5643 is further resolved into two peaks of CO(2--1) with an asymmetric
structure. The rest-frame 232\,GHz continuum, which likely pinpoints the
location of the AGN\footnote{The coordinates agree with those of the
  H$_2$O maser within their reported uncertainties $\pm 0.2\arcsec$ 
  \citep{Greenhill2003}.}, is located in between the two CO(2--1) 
peaks. This kind of lopsided gas morphology is the  characteristic
$m=1$ modes predicted by numerical simulations of gas dynamics on scales of
a few parsecs from an accreting super massive black hole and are due to 
angular momentum transfer \citep{HopkinsQuataert2010}.   However,
  we cannot rule out the possibility that the two CO(2--1) peaks
  reflect different gas excitation conditions in the nuclear region.

As can be see from Figure~\ref{fig:ALMAsmallFoV}, the 232\,GHz
  (1.3\,mm) continuum emission is produced by a compact source,
  although some resolved 
  emission seems to be present. To quantify this, we modeled the continuum in the uv plane using the UVMULTIFIT library
  \citep{MartiVidal2014} with a Gaussian function. The best fit
  produced a continuum flux of $1.8\pm0.2\,$mJy, with a deconvolved size (FWHM)
  along the major axis of $0.20\pm0.04$\arcsec \, and a size along the
  minor axis of $0.11\pm0.05$\arcsec  \, at $PA=210\pm 20$\arcdeg.  The
  orientation of the continuum extended emission would appear to
  follow the orientation of the inner contours of the  MUSE [O\,{\sc iii}] emission. It could
  thus be interpreted as either polar dust emission or synchrotron
  emission or both. We note  that radio emission in this
  direction is however observed both to the east and west of the AGN position 
\citep{Morris1985,
  Leipski2006}. Unfortunately the angular resolution of these radio
observations ($\sim 1-1.5\arcsec$) does not allow us to make any
further comparisons with the ALMA continuum emission.

\begin{figure*}

%\vspace{-0.1cm}
  
%\hspace{-0.7cm}
\includegraphics[width=0.34\textwidth]{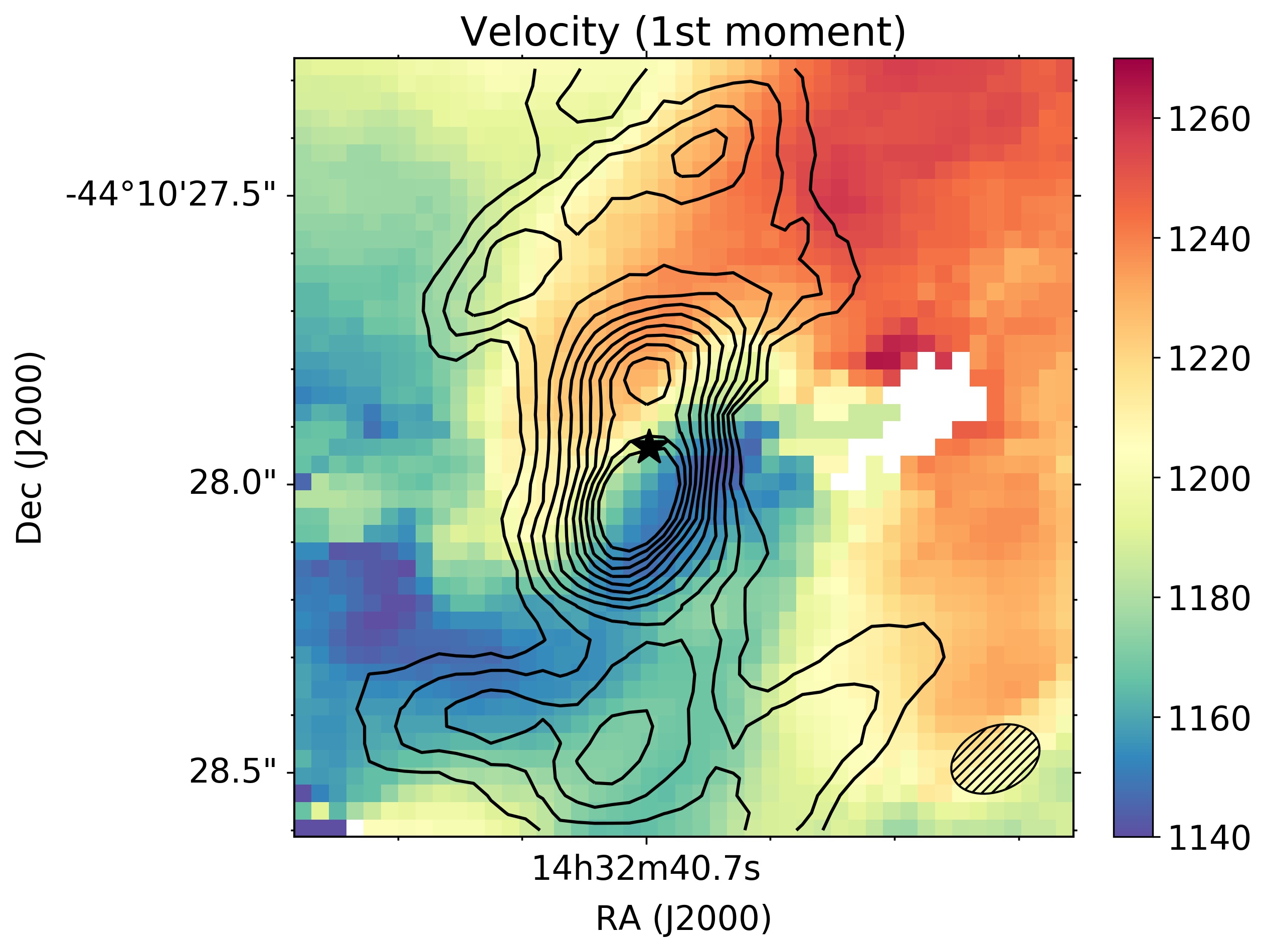}
%\hspace{-0.8cm}
\includegraphics[width=0.34\textwidth]{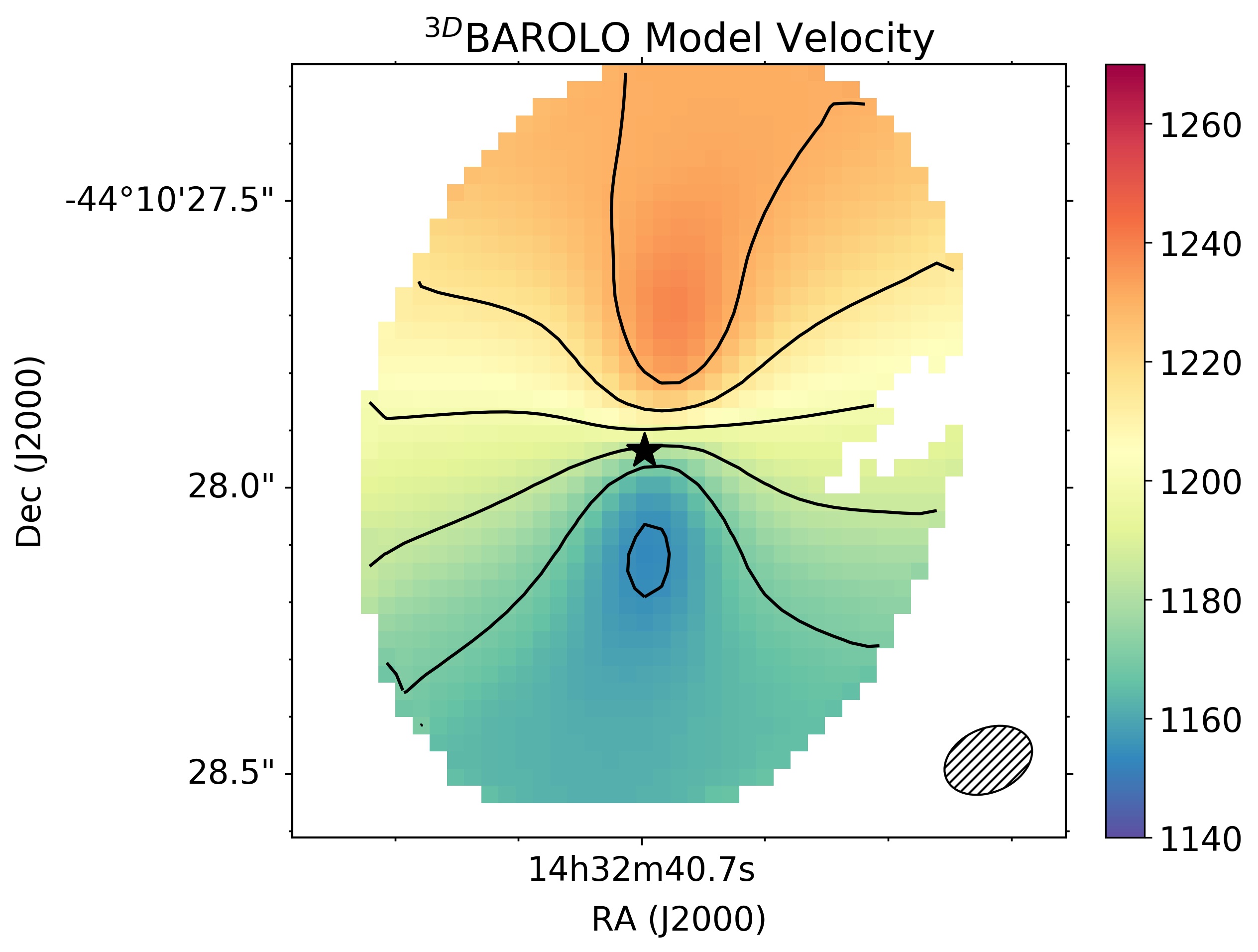}
%\hspace{-0.8cm}
\includegraphics[width=0.34\textwidth]{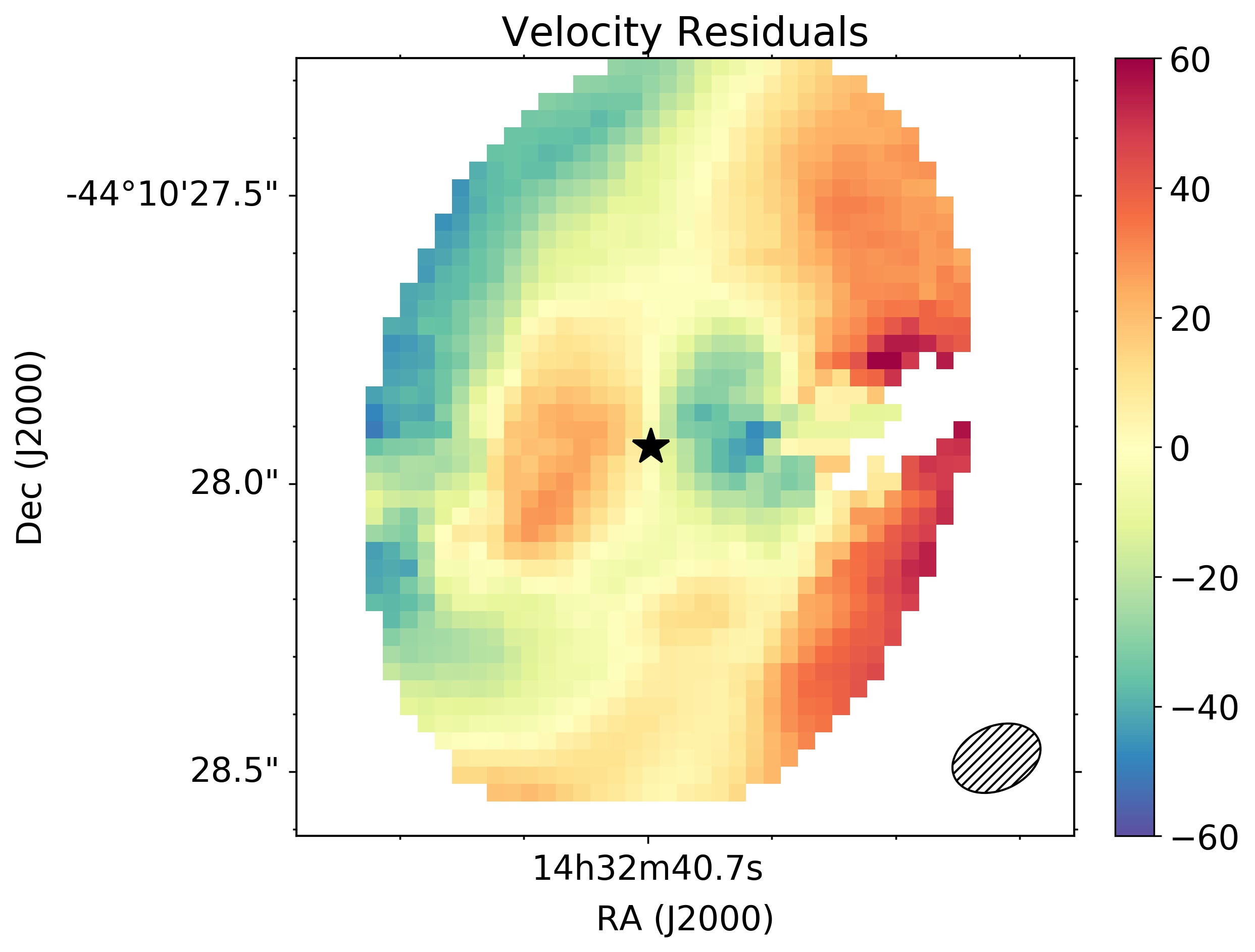}
%\vspace{-0.3cm}
\caption{Left panel. $^{\rm 3D}$BAROLO velocity field (1st
  moment map) from the $b=-0.5$ data cube. The black contours are
  CO(2--1) integrated line intensity. Middle panel:  
$^{\rm 3D}$BAROLO model of the velocity field without a radial
velocity component. The contours are the isovelocities of the model. Right panel: residual velocity field resulting
from subtracting the model from the observed velocity field.
 In all three panels the
  FoV and angular resolution are as in Figure~\ref{fig:ALMAsmallFoV} and the
  star symbol marks the peak of the rest-frame 232\,GHz continuum. The
vertical color bars indicate velocities in units of km\,s$^{-1}$.}\label{fig:BAROLOsmallFoV}
\end{figure*}

\begin{figure*}
  \hspace{2cm}
\includegraphics[width=0.34\textwidth]{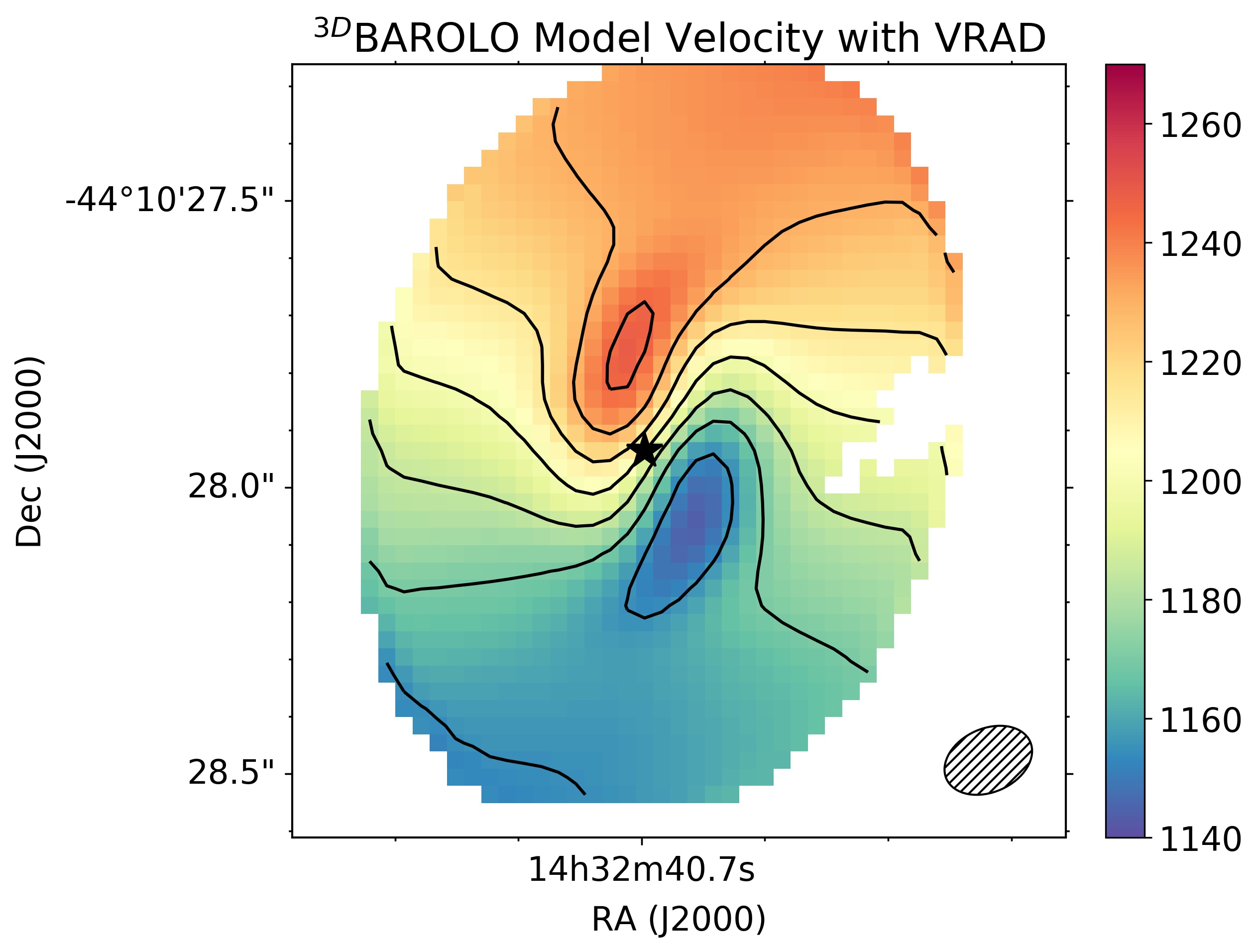}
\includegraphics[width=0.34\textwidth]{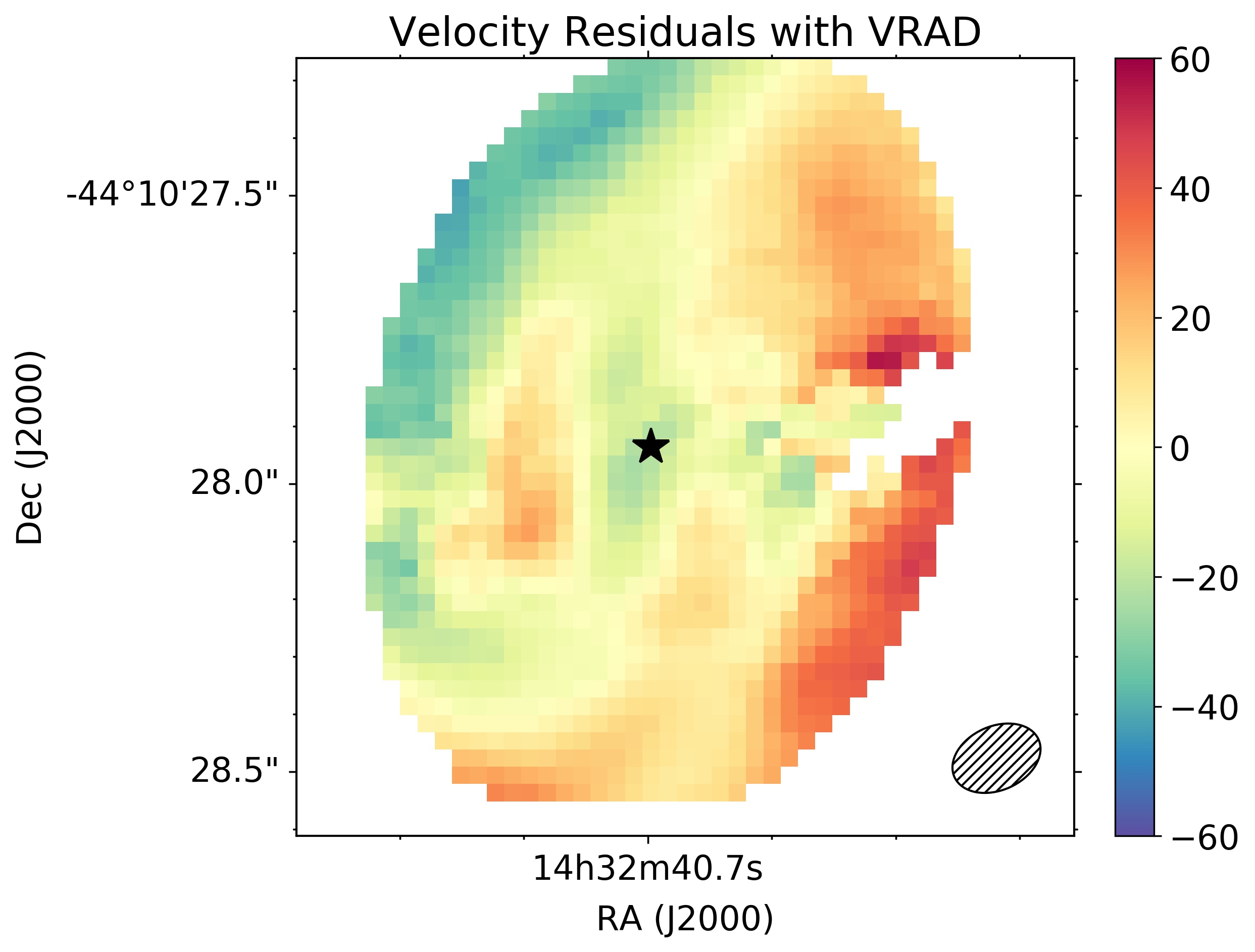}
\caption{The two panels are the same as the middle and right panels of
  Figure~\ref{fig:BAROLOsmallFoV} but the $^{\rm 3D}$BAROLO velocity
  field model includes a radial velocity component.}\label{fig:BAROLOsmallFoV_VRAD}
\end{figure*}

\subsection{Kinematics}\label{sec:kinemnuclear}

Figure~\ref{fig:BAROLOsmallFoV} (left panel) shows the mean velocity field (1st moment
map derived with $^{\rm 3D}$BAROLO) of the
CO(2--1) molecular gas in the nuclear region. A comparison with the larger scale CO(2--1)
kinematics shown in Figure~\ref{fig:BAROLO8x8} suggests that the nuclear disk might be
 tilted with respect to the host galaxy disk. We used again $^{\rm 3D}$BAROLO to construct
a disk model to fit the
kinematics of the nuclear region.  We first fitted a simple isophotal
model with the {\sc iraf} {\it ellipse} task to the nuclear
  CO(2--1) emission (see Figure~\ref{fig:ALMAsmallFoV}, right panel). The aim is 
  to provide $^{\rm
  3D}$BAROLO with initial values of the nuclear disk PA and 
inclination. From the outer isophotes
of the nuclear disk  (central $\sim 0.6\arcsec$) we can see
that the nuclear disk is almost in a north-south orientation (${\rm PA} \sim 354\arcdeg$) and the
measured ellipticity ($\epsilon =0.6$) implies a relatively high inclination.

We followed the same procedure as explained in
Section~\ref{sec:kinem8x8} to model the kinematics.
For the nuclear disk we fixed the kinematic center to the position of the
continuum peak. To derive the final $^{\rm 3D}$BAROLO  model we fixed
the nuclear disk geometry to ${\rm PA}=355\arcdeg$ and
$i=60\arcdeg$, and the systemic velocity to $v_{\rm sys}=1198\,$km 
s$^{-1}$. These were the average values derived when running 
$^{\rm 3D}$BAROLO leaving these parameters as well as the circular velocity and
velocity dispersion free. We note that the  inclination of the nuclear disk is
in good agreement with the inclination of ionization cone
derived by \cite{Fischer2013} from the modelling of the narrow line
region (NLR) kinematics, $i_{\rm cone} = 25\arcdeg$. If the cone is
perpendicular to the collimating nuclear disk then this is equivalent
to a nuclear disk inclination of 65$\arcdeg$. Finally, the
detection of a H$_2$O megamaser \citep{Greenhill2003} implies a highly
inclined view to the AGN.

We show the $^{\rm 3D}$BAROLO model and the residual mean-velocity field in
Figure~\ref{fig:BAROLOsmallFoV} (middle and right panels).
Since the nuclear disk appears to be seen almost edge-on we should not see velocity residuals
along the minor axis of the nuclear disk if the velocity field was
perfectly circular. The residuals are thus indicative of the presence of
non-circular motions in the nuclear disk.  As can be seen from this
figure, near the AGN peak there are redshifted residuals to the east
 and blueshifted residuals to the west. Outside the nuclear disk the residuals are related to the
velocity field of the larger scale disk of the galaxy which has a different
inclination and position angle, as discussed in Section~\ref{sec:kinem8x8}. 

To account for the non-circular motions seen in the nuclear disk,
we rerun $^{\rm 3D}$BAROLO including  a radial velocity component 
with the same assumptions for the rotating disk as above. An
inspection of 
Figure~\ref{fig:BAROLOsmallFoV_VRAD} reveals that the model with the extra
radial velocity component reproduces better the
observed velocity field with the residuals in the region of the
nuclear 
disk being of the order of $\pm 30\,{\rm km\,s}^{-1}$ or less. 
  The typical axisymmetric value of the radial velocity fitted by $^{\rm 3D}$BAROLO is of the
  order of 70\,km\,s$^{-1}$.

\begin{figure*}
\epsscale{0.9}
\plotone{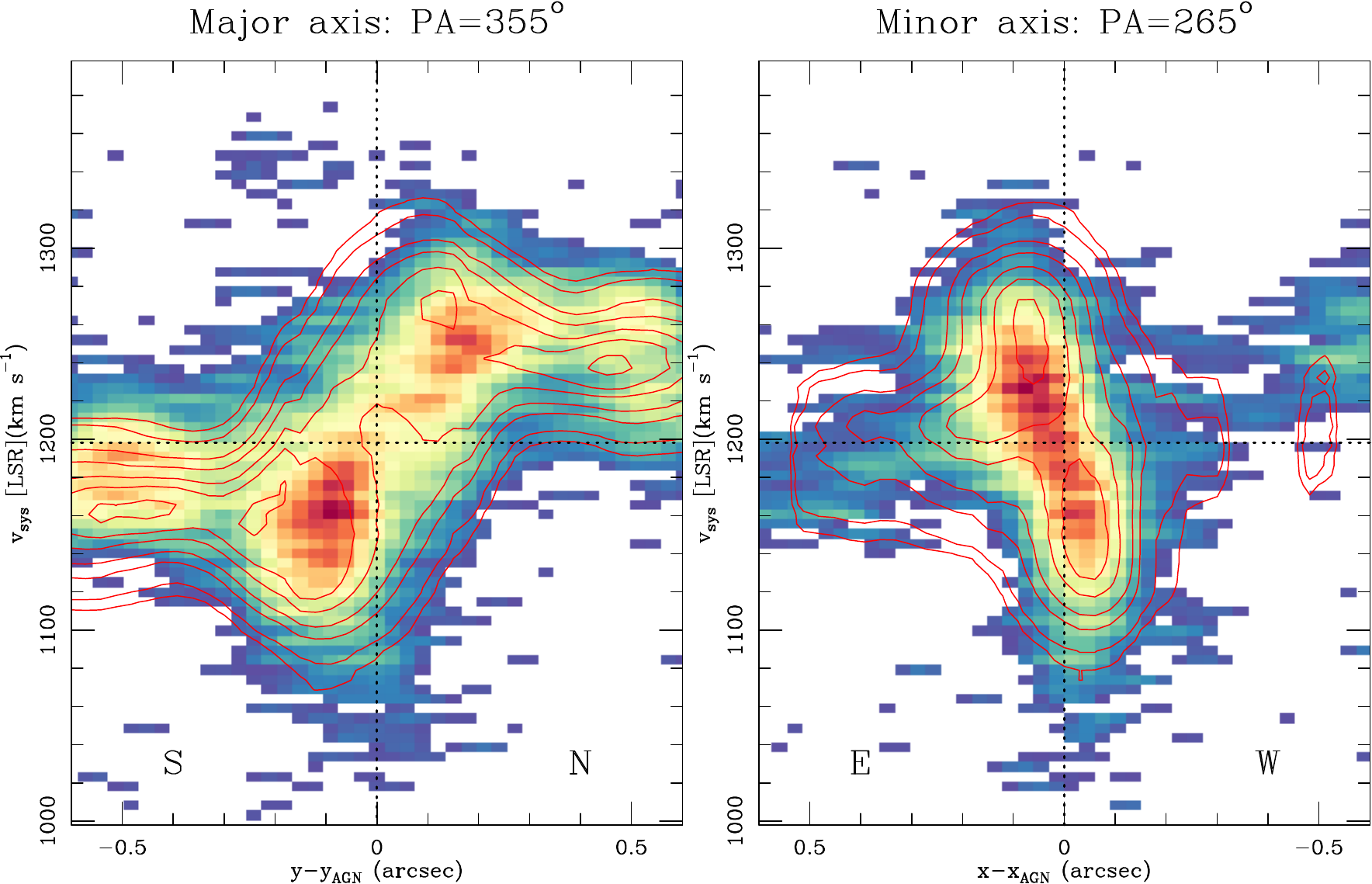}
\caption{In color are the position-velocity diagrams of the nuclear region extracted with
  an aperture size of 0.2\arcsec \, along the
  kinematic major and minor axes determined with $^{\rm 3D}$BAROLO for
  the nuclear region (see Section~\ref{sec:kinemnuclear}). The
horizontal lines are the systemic velocity fitted in the nuclear region and the vertical line
represents the position of the AGN as traced by the 232\,GHz continuum
peak. The red contours are the p-v diagrams derived from the $^{\rm 3D}$BAROLO model with
a  radial  velocity component (see Figure~\ref{fig:BAROLOsmallFoV_VRAD} and
text) at  1, 2.5, 10, 20, 40, 60 and 90\% of the peak value.}\label{fig:p-vdiagrams}
\end{figure*}

We also constructed  position-velocity (p-v) diagrams taken along the kinematic major
and minor axes
of the nuclear CO(2--1) disk oriented, according to the fits obtained
by  $^{\rm 3D}$BAROLO along ${\rm PA}^{\rm major}=355\arcdeg$ and
${\rm PA}^{\rm minor}=265\arcdeg$, respectively. We used an aperture  size of 0.2\arcsec \, which
is approximately equal to the beam size of the $b=-0.5$ data cube.
As seen from
Figure~\ref{fig:p-vdiagrams}, the CO(2--1) emission in
the nuclear disk is spread over a wide range of velocities more or
less  symmetrically around $v_{\rm sys}=1198$\,km~s$^{-1}$: $v-v_{\rm
  sys}\simeq[-160,\, +130]$\,km~s$^{-1}$. The two p-v diagrams
illustrate the kinematic decoupling of the nuclear CO(2--1) disk relative to the larger scale disk of the galaxy. The amplitude and
even the sign of the characteristic mean  line-of-sight velocities of the emission show a significant change
beyond $r\simeq0.3-0.4\arcsec$ \, (30~pc) and
$r\simeq0.1-0.2\arcsec$ \, (14~pc) in the major and minor axis p-v
diagrams, respectively. These physical scales coincide with those
derived from the morphology of the CO(2--1) intensity (Figure~\ref{fig:ALMAsmallFoV}).

The behavior seen in the p-v diagram identifies the transition from the nuclear disk to the larger
scale disk modeled in Section~\ref{sec:kinem8x8}. 
Furthermore, the CO(2--1) nuclear disk shows strong non-circular
motions. 
An inspection of the minor-axis p-v diagram indicates that
 the maximum observed deviations from rotational motions reach $\pm 100 \,{\rm km\, s}^{-1}$. The emission
appears blueshifted (redshifted) on the
western (eastern) side of the  nuclear disk, as also seen in the mean velocity
residual map in Figure~\ref{fig:BAROLOsmallFoV}. 
In Figure~\ref{fig:p-vdiagrams}, we also plotted as contours
  the p-v diagram resulting from the $^{\rm 3D}$BAROLO fit with the
  radial velocity component. It is clear that this model is able to
  reproduce fairly well the non-circular motions in the nuclear
  region. We
discuss the possible interpretations of this radial velocity component
in Section~\ref{sec:conclusions_nuclear}.

\subsection{Molecular gas mass} 

We measured a CO(2--1) line intensity over the nuclear region
approximately of $20\,$Jy km s$^{-1}$ over a region of $\sim 0.4\arcsec
\times 0.6\arcsec$. Assuming a
CO(1--0)/CO(2--1) brightness temperature ratio of one
and using the relation of \cite{Sakamoto1999} 
with a Galactic CO-to-H$_2$ conversion factor of $X=2\times 10^{20}\,{\rm
  cm}^{-2} $ (K km s$^{-1})^{-1}$ we derived a molecular gas mass in
the nuclear disk of $M({\rm H}_2)=1.1\times 10^7\,M_\odot$.
The typical uncertainty for the X-factor is 2--3 for gas clouds in
our Galaxy \citep{Bolatto2013}. The nuclear disk of NGC~5643 is
about a factor of 10  more massive than the torus in NGC~1068
\citep{GarciaBurillo2016, Imanishi2018}.  This is not surprising given
the different sizes measured in
CO(2--1) for NGC~5643 (see Section~\ref{sec:morphnuclear}) and
in  NGC~1068 using dense molecular gas tracers
  \citep{GarciaBurillo2016, Imanishi2018}.
However, the size of the torus of NGC~1068 measured in the CO(2--1) and CO(3--2)
transitions is about 20\,pc and thus comparable to NGC~5643 (Garc\'{\i}a-Burillo et al. 2018,
in preparation).
On the other hand, the derived molecular gas mass of the nuclear disk in NGC~5643 agrees  well
with the average gas mass  for a sample of
Seyfert galaxies from dynamical constraints assuming a gas fraction of 10\%
\citep{Hicks2009}.

We also estimated the dynamical mass enclosed in the nuclear disk
  regions by assuming a spheroid, $M_{\rm dyn}(R) = 2.32\times10^5\,R\,V^2(R)$
  in $M_\odot$ 
  with R expressed in kpc, V in km\,s$^{-1}$ 
\citep{Lequeux1983}. For
  the velocity we took the terminal value of $v_{\rm term} = 100\,{\rm
    km\,s}^{-1} /\sin(i) = 115\,$km\,s$^{-1}$ at $r=0.1-0.2$\arcsec \,
  (see
  Figure~\ref{fig:p-vdiagrams}) which is more appropriate for inclined
  disks. This leads to a dynamical mass of approximately $5\times
  10^7\,M_\odot$, which would imply a gas fraction of approximately 20\%.

The high
  molecular gas surface density in the nuclear disk of NGC~5643
  would imply some star formation activity in the nuclear region as found
  for other Seyferts \citep{Hicks2009}. 
The detection of  emission from the
$11.3\,\mu$m polycyclic aromatic hydrocarbon (PAH) feature in the  central
$0.4-0.7\arcsec \sim 33-57 \,$pc \citep[see ][]{Hoenig2010,
  GonzalezMartin2013} might be indicating on-going or recent star
formation activity in the nuclear region. 

\subsection{Column Density}

Having the molecular gas mass in the disk and the approximate size we
computed an average column density of
$N ({\rm H}_2)\sim 4 \times 10^{23}\,{\rm cm}^{-2}$. This value is the
column density averaged over the nuclear disk size of $\sim 0.4\arcsec
\times 0.6\arcsec$. The averaged column density is
similar to those  derived for Seyfert galaxies
by \cite{Hicks2009} on similar physical scales using the near-infrared
$2.12\,\mu$ H$_2$ line. If we only focus on
the AGN position, as given by the 232\,GHz continuum peak (see
Figure~\ref{fig:ALMAsmallFoV}) and at our angular resolution, we would
derive a column density of  $N({\rm H}_2)\sim 5 \times 10^{23}\,{\rm cm}^{-2}$. 
At the peak of the CO(2--1) emission the value of the column density
is  $N({\rm H}_2)\sim 7\times 10^{23}\,{\rm cm}^{-2}$.

The derived H$_2$ column density is short of the value
derived from X-ray observations which provides
 a lower limit for the obscuring column
density toward the AGN in NGC~5643 of $N_{\rm H} ({\rm los})>5 \times 10^{24}\,{\rm
  cm}^{-2}$ \citep{Annuar2015}. However, the X-ray
  column is modelled through a narrow pencil beam towards the
  accretion disk. For type 2 AGN one would expect it to intersect high
  column clouds in the torus (or even
the broad line region). On the other hand, the column measured from
our  ALMA data is averaged over a larger area which is defined by the beam of $\sim 13\,{\rm
  pc} \times 9\,{\rm pc}$ of the observations. 

Another source of uncertainty in this
comparison is the assumed CO-to-H$_2$ conversion factor.
\cite{Wada2018} computed the
CO line chemistry from their radiation-driven
fountain model fitted to the Circinus galaxy to derive the CO-to-H$_2$ conversion factor. They
demonstrated that this conversion factor depends strongly on the integrated intensity of
the line for a given line of sight. 
For the brightness temperature derived at the location of the AGN or the peak of the CO(2--1) emission
of NGC~5643 the predictions for CO-to-H$_2$ conversion factor for the
CO(2--1) line are 5-10 higher than the Galactic value
with a large scatter (see their figure~5). We note that the inclinations of NGC~5643 and Circinus
are similar ($i\sim 60-65$\arcdeg \, vs. $i\sim75$\arcdeg). We
  note, however, that empirically  the CO conversion factor tends to
  be lower in the centers of galaxies \citep{Sandstrom2013}.

\begin{figure*}
%\epsscale{1.1}

\includegraphics[width=0.99\textwidth, angle=0]{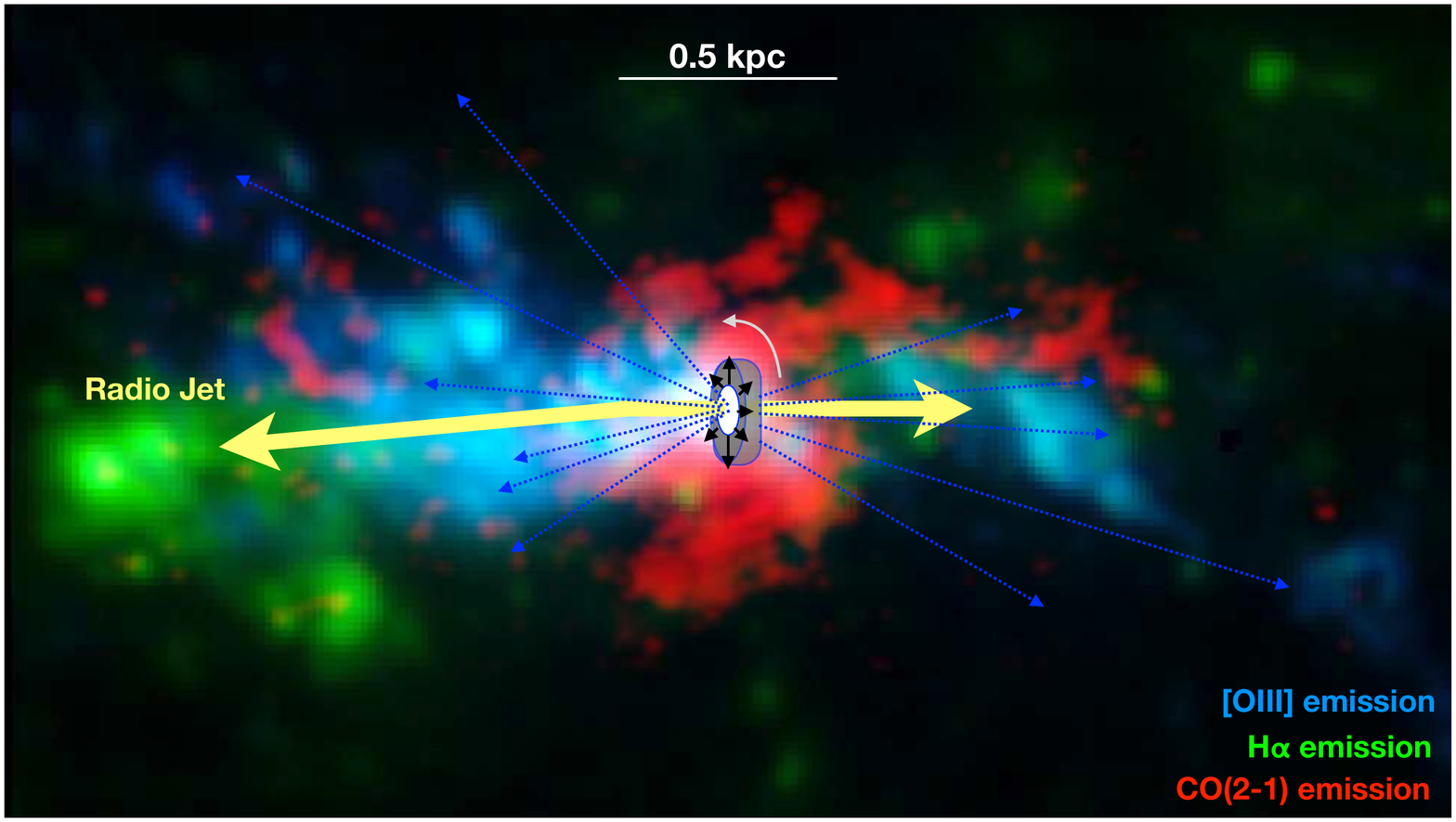}
%\plotone{figure9.ps}
\vspace{-1cm}
\caption{Cartoon showing the nuclear and circumnuclear region of 
NGC~5643 with a FoV similar to that of the lower panel of Figure~1. The RGB image was 
constructed with the MUSE [O\,{\sc iii}]
image (blue), the MUSE H$\alpha$ image (green) and the ALMA CO(2--1)
image (red). The last image was smoothed with a Gaussian
function with FWHM$\sim 0.4$\arcsec. The nuclear
disk/torus and the radio jet are not drawn to scale and the physical
scale bar is also approximate.}\label{fig:cartoon}
\end{figure*}

\section{Discussion and Conclusions}\label{sec:conclusions}

\subsection{Circumnuclear scales}
To further understand the connection between the cold molecular gas in
the circumnuclear region of NGC~5643 and the outflow detected in
ionized gas we produced a red-green-blue (RGB) image using the MUSE [O\,{\sc iii}]
image in blue, the MUSE H$\alpha$ image in green (see Section~\ref{sec:MUSEobs}), and the ALMA CO(2--1)
image in red. The result is shown in Figure~\ref{fig:cartoon} with a
FoV approximately equal to that of the lower panel of Figure~\ref{fig:ALMAfullFoV}. The RGB image shows 
nicely the ionization cone as traced by the [O\,{\sc iii}] emission on the east
side of the galaxy. Close to the AGN position the [O\,{\sc iii}] emission from
the counter-cone to the
west is obscured by material in the host galaxy which is traced by the
CO(2--1) nuclear spiral. The H$\alpha$ emission on this side of the galaxy is less
obscured, as
expected \citep[see][for an in-depth discussion]{Cresci2015}. 

\cite{Fischer2013} interpreted the extended [O\,{\sc iii}] emission
observed in some Seyfert galaxies as the result  of the intersection between the disk of
the host galaxy and the bicone. From their modeling of the NLR
kinematics  of NGC~5643 they inferred an
angle between the NLR bicone axis and the normal to the host
galaxy disk of $42\arcdeg$. We derived an angle difference between the
normals of the 
galaxy disk and the nuclear disk fitted with $^{3D}$BAROLO of
$25\arcdeg$ ($30\arcdeg$ if we
take the value they used for the inclination of the galaxy). They also
derived a cone half outer opening angle of $\theta_{\rm
  max}=55\arcdeg$. This  implies that this rather wide ionization cone
would intercept the host galaxy but only to the southeast of the
AGN where the radio jet is also impacting on the host galaxy and
producing positive feedback \citep{Cresci2015}.

To the northeast of the galaxy nucleus, some of the
[O\,{\sc iii}] filaments are coincident with the
regions  with CO(2--1) with redshifted
velocity components
(see Figures~\ref{fig:ALMAOIIIcolormap} and ~\ref{fig:lineprofiles}) and are
mostly along the minor axis of the host galaxy. As interpreted for the
hot molecular gas H$_2$ emission by \cite{Davies2014}, these components are probably due to gas
excited on the edge of the cone. The spiral morphology of the CO(2--1)
emission suggests that this emission is in the disk of the galaxy, and so is the
gas with the redshifted velocity component.  With this geometry the non-circular motions
are due to molecular gas  being
pushed outwards in the disk of the galaxy (note that northeast is the
  far side of the galaxy disk and southwest the near side) by the AGN wind near the {\it fourth} spiral arm in the
circumnuclear region. The region with the blueshifted velocity component to the southwest of the
AGN can also be explained as gas being pushed outwards by the ionized outflow and moving
towards us. 
%In this scenario the opening angle of the ionization cone needs to be
%larger than that modeled by \cite{Fischer2013}
%in order to be able to intercept the disk of the galaxy to the
%northeast as well as the southeast of the AGN.
%The interaction between the radio jet and the cold molecular gas
%clouds traced by CO(2--1) in
%IC~5063 \citep{Morganti2015}. We note, however, that  IC~5063 is a
%radio bright Seyfert galaxy and thus the radio jet is more energetic
%than in NGC~5643, and thus the gas velocities in IC~5063 are higher.
A similar interaction between the outflow detected in ionized gas and the
cold molecular gas
in the galaxy disk with blueshifted and redshifted 
velocity components has been observed in the Circinus galaxy \citep{Zschaechner2016}.

We also plotted in the cartoon of Figure~\ref{fig:cartoon} the approximate
orientation of the radio emission/jet \citep{Morris1985,
  Leipski2006} pointing to the location of the H\,{\sc ii} regions
with bright H$\alpha$ emission. \cite{Cresci2015} proposed that these
regions are the result of positive feedback induced by gas compression
by the AGN outflow. Although faint, we detected CO(2--1) emission at
the location of these star-forming clumps. 
The passage of the radio jet through the disk of the 
galaxy might be responsible for clearing the
CO(2--1) molecular gas in  the eastern side spiral arm in a region
$\sim 5\arcsec$ the nucleus
of NGC~5643 (see Figure~\ref{fig:ALMAfullFoV}).

\subsection{Nuclear scales}\label{sec:conclusions_nuclear}
We have detected a massive ($M({\rm H}_2)=1.1\times 10^7\,M_\odot$)
nuclear disk in NGC~5643. The size of the nuclear disk measured from the Band 6 CO(2--1)
integrated emission is approximately 26\,pc (FWHM). This is in fairly good agreement with the 
typical sizes (radius of 30\,pc) derived for nuclear disks from hot molecular gas
H$_2$ $2.12\,\mu$m by \cite{Hicks2009}. 
The disk/torus of NGC~5643 is a factor of two-to-three larger than that detected
by ALMA in CO(6--5) in the Seyfert 2 galaxy NGC~1068
\citep[$\sim 7-10\,$pc, see][]{GarciaBurillo2016}. One possible explanation is that the
CO(6--5) line traces denser gas than CO(2--1) and thus it probes
molecular gas that needs to be closer to the AGN to
be excited. Indeed, our new ALMA CO(2--1) and CO(3--2) observations of NGC~1068 reveal a
larger structure (diameter $\sim 20$\,pc) than seen in CO(6--5) which is also well connected
with the circumnuclear disk 
(Garc\'{\i}a-Burillo et al. 2018, in preparation). The molecular gas
column density averaged over the ALMA beam of our observations and at
the AGN location  is  $\sim 5 \times 10^{23}\,{\rm cm}^{-2}$.  The material in the nuclear
disk is likely responsible, at least in part, for obscuring the AGN in NGC~5643.

The modeling of the CO(2--1) kinematics in the nuclear region  clearly implies the
presence of non-circular motions, with a maximum amplitude of $100\,{\rm
    km\,s}^{-1} /\sin(i)$,  in the inner $r\sim 0.2\arcsec$
(16\,pc). However, their interpretation requires a
detailed knowledge of the nuclear potential and in particular whether there is a
nuclear bar in NGC~5643 or not. Using near-infrared imaging
observations, \cite{Jungwiert1997} detected an isophotal
twist between radial distances of $r=3\arcsec$ and
$r=30\arcsec$. This feature could be indicative
of the presence of a nuclear  bar with $PA=-49\arcdeg$. In fact, it is
possible that the nuclear arc structures could be associated with such a
bar although they are on smaller
scales (see Figure~\ref{fig:ALMAsmallFoV}).  In this scenario the
radial motions seen in the inner 1\arcsec \, could 
be interpreted as gas elliptical stream lines in a nuclear bar. In a dynamically
decoupled nuclear bar  the observed kinematics could be explained as both inflowing and outflowing radial motions.

In the absence of a putatively decoupled nuclear stellar bar,
the canonical gas response to the large scale bar in the framework of
the epicyclic approximation  inside the
inner Lindblad resonance would favor the presence of inward radial motions
\citep[see Figure~3
of][]{Wong2004}.
% and even in the NGC~5643-adapted model of
%\cite{Davies2014}.  
% This is not to say that the gas is
%losing angular momentum in this region. Rather, it is likely the
%opposite, gas is gaining angular momentum inside the ILR but the
%residuals of the velocity field suggest a predominance of inward radial motions.
Therefore, in a coplanar geometry with the nuclear disk  (see the
  assumed orientation of the nuclear disk in Figure~\ref{fig:cartoon}) the
non-circular motions are not due to the large scale bar and 
could  be  explained as due to material outflowing in the plane of the
nuclear
disk. This reproduces the blueshifted velocities on the western  side of the
nuclear disk and redshifted velocities on
  the eastern  side of the nuclear disk (see p-v along the minor
  axis of the nuclear disk in Fig.~\ref{fig:p-vdiagrams}). The ionized
gas in NGC~5643 is outflowing (blueshifted emission  at the  AGN location in the MUSE
[O\,{\sc iii}] emission) perpendicular to the nuclear disk
\citep{Cresci2015}. The different location for the molecular
gas and ionized gas is predicted by the radiation-driven fountain
model 
simulations done for the Circinus galaxy \citep{Wada2016,Wada2018}. In particular this model
shows that most of
the nuclear molecular gas emission is expected
in the equatorial plane of the nuclear disk with no dense {\it molecular
winds}  outflowing  along  the  rotational  axis, while the ionized
gas is outflowing perpendicular to the disk. A similar {\it
  equatorial} outflow has been observed in the Seyfert 2 galaxy
NGC~5929 by \cite{Riffel2014} on scales even larger ($\sim 300\,$pc)
than in NGC~ 5643.

We conclude that the Band 6 CO(2--1) observations have resolved a massive nuclear
molecular gas rotating disk/torus in NGC~5643 with strong non-circular motions. The inclination of the 
nuclear disk agrees with the inclination required to model the ionization cone
emission traced by [O\,{\sc iii}] and H$\alpha$ emission
\citep{Fischer2013}. Thus the CO(2-1) disk/torus is likely
collimating the ionization
cone of NGC~5643 as well as obscuring 
the AGN. In the framework of the AGN Unified Model \citep{Antonucci1993}, the detected
disk could be interpreted as the obscuring torus.

\acknowledgments

We thank Enrico di Teodoro for providing advice to
run $^{\rm 3D}$BAROLO and an anonymous referee for suggestions
  that helped improve the paper.

A.A.-H. acknowledges support from the Spanish Ministry of Economy and
Competitiveness through grant AYA2015-64346-C2-1-P which was party funded by the FEDER
program and from CSIC grant PIE201650E36.  A.A.-H., M.P.-S., and
A.B. acknowledge
support from the Royal Society International Exchange Scheme 
under project IE160174. A.L. acknowledges support from the Spanish
Ministry of Economy and Competitiveness through grant ESP2015-68964.  M.P.-S. and D.R.
acknowledge support from STFC through grant ST/N000919/1. 
T.D-S. acknowledges support from ALMA-CONICYT project 31130005 and
FONDECYT regular project 1151239. P.G. acknowledges support from STFC (ST/J003697/2). A.H.-C. acknowledges
funding by the Spanish Ministry of Economy and
Competitiveness under grant AYA2015- 63650-P.
S.F.H. acknowledges support from the EU/Horizon 2020 ERC Starting Grant DUST-IN-THE-WIND (ERC-2015-StG-677117).
C.R.A. acknowledges the Ram\'on y Cajal Program of the Spanish Ministry of Economy and Competitiveness through
project RYC-2014-15779 and AYA2016-76682-C3-2-P. 

This paper makes use of the following ALMA data:
ADS/JAO.ALMA\#2016.1.00254.S. ALMA is a partnership of ESO
(representing its member states), 
NSF (USA) and NINS (Japan), together with NRC (Canada), MOST and ASIAA
(Taiwan), and KASI (Republic of Korea), in cooperation with the
Republic of Chile. The Joint ALMA Observatory is operated by ESO,
AUI/NRAO and NAOJ. Based on observations collected at the European
Organisation for Astronomical Research in the Southern Hemisphere  
under ESO programme: 095.B-0532(A).


\vspace{5mm}
\facilities{ALMA, ESO (VLT/MUSE)}



\end{document}